\documentclass{aa}

\usepackage{color}
\usepackage{float}
\usepackage{graphicx}
\usepackage{txfonts}
\usepackage[normalem]{ulem}
\usepackage{natbib}
\bibpunct{(}{)}{;}{a}{}{,} 
\usepackage{amssymb}
\usepackage{amsfonts}
\usepackage{tabularx}
\usepackage{array}
\usepackage{booktabs}
\usepackage[fleqn]{amsmath}
\usepackage{subfigure}
\usepackage{diagbox}
\usepackage{siunitx}
\usepackage{hyperref}

\hypersetup{colorlinks = true, citecolor = {blue}, urlcolor= {blue}}

\begin{document} 

\nolinenumbers

\title{Origin of Ca~II emission around polluted white dwarfs}

\subtitle{}

\author{
V. Fröhlich
    \inst{1,2,3}\thanks{E-mail: \href{frohlich.viktoria@csfk.org}{frohlich.viktoria@csfk.org}}\orcid{0000-0003-3780-7185}
    \and
Zs. Regály
    \inst{1,2}\orcid{(0000-0001-5573-8190)}
}

\institute
{
{HUN-REN Konkoly Observatory, Research Centre for Astronomy and Earth Science, Konkoly-Thege Mikl\'os út 15-17, 1121, Budapest, Hungary}
\and
CSFK, MTA Centre of Excellence, Budapest, Konkoly Thege Miklós út 15-17., H-1121, Budapest, Hungary
\and
{E\"otv\"os Lor\'and University, P\'azm\'any P\'eter s\'et\'any 1/A, 1117 Budapest, Hungary}
}

\date{Received April 22, 2024}
 
\abstract
   {
   Dozens of white dwarfs with anomalous metal polluted atmospheres are currently known to host dust and gas discs.
   The line profiles of the Ca~II triplet emitted by the gas discs show a significant asymmetry.
   In recent decades, researchers have also discovered several minor planets orbiting such white dwarf stars. 
   }
   {
    The most challenging burden of modelling gas discs around metal polluted white dwarfs is to simultaneously explain the asymmetry and metal pollution of the star's atmosphere over a certain period of time.
    Furthermore, models should also be consistent with other aspects of the observations, like the morphology of the emission lines.
    This paper aims to construct a self-consistent model to explain the simultaneous white dwarf pollution and Ca~II line asymmetry over at least three years.
   }
   {
    In our model an asteroid disintegrates in an eccentric orbit, periodically entering below the star's Roche limit.
    The debris resulting from the disintegration sublimates at a temperature of 1500~K, producing gas that viscously spreads to form a disc. 
    The evolution of the discs is studied over a period of 1.2 years (over 21000 orbits) using two-dimensional hydrodynamic simulations. 
    Synthetic Ca~II line profiles are calculated using the surface mass density and velocity distributions provided by the simulations, for the first time taking into account the asymmetric velocity distribution in the discs.
    }
   {
    An asteroid disintegrating on an eccentric orbit gives rise to the formation of an asymmetric disc and  asymmetric Ca~II triplet emission.
    Our model can explain the periodic reversal of the redshifted and blueshifted peak of the Ca~II lines caused by the precession of the disc on timescales of 10.6 to 177.4 days.
   }
   {
   In summary, our work suggests that the persistence of Ca~II asymmetry over decades and its periodic change in the peaks can be explained by asteroids on eccentric orbits in two scenarios.
    In the first case, the asteroid disrupts on a short timescale (couple of orbits), and the gas has a low viscosity range ($0.001<\alpha<0.05$) to maintain the Ca~II signal for decades. 
    In the other scenario, the asteroid disrupts on a a timescale of a year, and the viscosity of the gas is required to be high, $\alpha=0.1$.
   }


\maketitle

\nolinenumbers

\section{Introduction}
\label{sec:intro}

Despite the strong gravitational acceleration at their surface, 25-50 per cent of galactic white dwarfs (WDs) show evidence of metal pollution \citep{zuckermanetal03, koesteretal14}.
The effective temperature of these metal-polluted, so-called DZ type WDs ranges from 5.000 to 23.000 K, implying a cooling age of 40 million to 2 billion years \citep{hollandsetal17, koesteretal14}.
This is at least an order of magnitude higher than the settling times of the pollutants, which ranges from days to a few million years \citep{paquetteetal86, metzgeretal12}.
Metal pollution has been observed to persist for at least twelve years \citep{gaensickeetal08, hartmannetal14, hartmannetal16, manseretal16a}, and the fading of the phenomenon is yet to be reported.

Most studies assume that interplanetary matter, such as planets, moons, asteroids, comets, dust and gas, is the source of metal pollution \citep{zuckermanetal03, dufouretal07, zuckermanetal10, kleinmanetal13, koesteretal14, gentilefusilloetal15, kepleretal15, kepleretal16}.
The amount of accreted material can range from the mass of Phobos to that of Pluto (refer to Fig.~6 of \citealp{veras16}).

The idea of interplanetary matter polluting WDs is supported by the fact that rocky minor planets have been discovered around DZ WDs.
The first of such bodies was found in the WD~1856+534 system along with five additional small bodies \citep{crolletal17, gaensickeetal16, rappaportetal16, vanderburgetal15, xuetal16}.
All of these minor planets orbit the WD on circular and coplanar orbits with periods of 4.5 to 4.9~hr (corresponding to semi-major axes of $5\times 10^{-3}$ to $5.6\times 10^{-3}~\mathrm{au}$) and have radii of 1 to 100 km.

Minor planets on such tight orbits have certainly crossed the Roche radius of the WD and have thus begun to shed mass.
Other, newly discovered rocky planets around DZ WDs also have in common an orbit that lies at or below the Roche limit \citep{manseretal19, vanderboschetal20, guidryetal21}.
Furthermore, the observation of asymmetric transit light curves strongly suggests a cometary tail of volatiles around these objects.
Both observations imply that terrestrial planets are being seen during the process disruption around DZ WDs.
Note that gas giants of 2.5 to 14 Jupiter mass have also been discovered around WDs, however, not all of their stars are metal polluted \citep{luhmanetal11, gaensickeetal19, vanderburgetal20, mullallyetal24}.

Over 50 DZ WDs have been observed to host dust discs \citep{bergforsetal14}, 21 of which also host a gaseous counterpart \citep{gaensickeetal06, gaensickeetal07, gaensickeetal08, gaensicke11, farihietal12, melisetal12, wilsonetal14, dennihyetal20, gentilefusilloetal21, melisetal20}.
Currently, all DZ WDs that are known to host a gas disc also share a dusty component.
However, many WDs have been observed that are only accompanied by a dust disc.
Emphasise that both dust and gas discs are present in the systems that host rocky minor planets.
The stability of discs during the giant phase of the host star is not possible due to strong stellar winds, thus discs must have formed after the star had become a WD \citep{veras16}.

It has been observed that both the dust and the gas discs extend below the Roche radius of the WD.
Dust discs are typically observed to lie between $0.6~R_\odot-1.2~R_\odot$ \citep{farihi16}, with a mass ranging from $10^{19}-10^{24}~\mathrm{g}$ \citep{manseretal16a}. 
Gas discs, on the other hand, are found at $0.5~R_\odot-1.2~R_\odot$ \citep{gaensickeetal08, farihi16} and have a mass ranging from $10^{14}-10^{19}~\mathrm{g}$ \citep{hartmannetal16}.
The inner boundary of the gas discs should coincide with a temperature of 5000~K, since no disc emission lines are observed in the far-ultraviolet region of the spectra \citep{hartmannetal16}.

Gas discs are typically discovered through the infrared emission of the Ca~II triplet.
If the discs were Keplerian, the emission lines would show symmetrical double peaks \citep{hornemarsh86}.
However, this is not the case for the gas discs of DZ WDs, which show an asymmetry in the intensity of the red- and blueshifted peaks.
This is likely due to a non-Keplerian density or velocity distribution in the disc.
The asymmetry has been observed to persist for as long as twelve years \citep{wilsonetal14, manseretal16a, manseretal16b, cauleyetal18, dennihyetal18, dennihyetal20}, but its disappearance has also been observed \citep{gaensickeetal08, hartmannetal14}.

Additionally, a periodic fluctuation in the asymmetry of the Ca~II lines has been detected.
This is hypothesised to be caused by the precession of an eccentric disc (e.g. \citealp{cauleyetal18, gaensickeetal08, hartmannetal14, manseretal16a}).
The period of the change in asymmetry ranges from 123 minutes to 37 years \citep{hartmannetal16, manseretal19}.
However, to accurately map these changes, it is necessary to measure at a cadence that is a fraction of the precession period (great examples are \citealp{dennihyetal18, manseretal19}).

The fact that the WDs at the heart of these systems are metal polluted means that the material of the discs is lost to the WD through accretion.
Taking into account the time scale on which anomalous metallicity has been observed and the settling times of the pollutants, one must assume a continuous replenishment of the disc material \citep{koester09}.
This rules out the direct infall of a planet, asteroid, moon or comet onto the star \citep{debesetal12ast, verasetal14, verasetal16, payneetal16, bhatt85} as a pollution mechanism.
Furthermore, the above scenario is unable to explain the origin of the discs around DZ WDs.

On the other hand, continuous replenishment requires the presence of a constantly disturbed asteroid belt \citep{bonsoretal11, debesetal12ast, frewenhansen14, verasetal16, nixonetal20} or a disintegrating planet, asteroid or comet \citep{jura03, jura08, debesetal12giant, verasetal13}.
The latter scenario is considered to be the most likely origin of the dust discs \citep{debessigurdsson02, jura03, malamudperets20b, malamudperets20a, trevascusetal21, verasetal14}.

Numerical simulations suggest that bodies with semi-major axes of a few au can migrate to the Roche limit due to tidal dissipation or perturbations from other planets in the system, and that terrestrial planets are more likely to migrate inwards than gas giants \citep{verasetal16, oconnorlai20}.
This, together with the high density and small size of the planets observed around DZ WDs and the spectra of the contaminated stars, strongly supports the hypothesis that Earth-like bodies are being disrupted around DZ WDs.
The observed discs are devoid of hydrogen \citep{hartmannetal11}, which is also consistent with the above scenario.

With regard to the origin of the gas discs, some studies assume that the disintegrating body contains intrinsic gas, so it is produced immediately as the body passes the Roche limit \citep{trevascusetal21, fortinarchambaultetal20}.
Others investigate the effect of a primordial dust or gas disc, which is later perturbed by a disrupting body \citep{farihietal18, kenyonbromley17b, malamudetal21, swanetal20}.
\citet{rafikov11} and \citet{metzgeretal12} assume that the gas is produced at a specific sublimation radius:
dust particles are pulled into tighter orbits by the radiation of the host star (Poynting-Robertson drag), where they sublimate due to the higher temperatures.
The authors modelled the formation of dust and gas discs simultaneously. 
However, they only considered disc formation in one dimension.

Multiple studies have found that in order for the disc morphology to be consistent with the observed structures, the disrupting body must have an eccentric orbit \citep{rafikov11, metzgeretal12, kenyonbromley17b, fortinarchambaultetal20, trevascusetal21}. 
As such, several studies have modelled a series of eccentric, precessing gas rings and have calculated synthetic Ca~II emission lines using radiative transfer models \citep{gaensickeetal06, gaensickeetal08}.
Some have attempted to fit the observations with Monte Carlo methods \citep{goksuetal23} or have assumed gas discs that are not in local thermodynamic equilibrium \citep{hartmannetal14, hartmannetal16}.

Our aim is to develop a model that can simultaneously explain the anomalous metal pollution of WDs and the asymmetric Ca~II lines over at least a 1.2-year time span, while being consistent with the observations regarding line morphology and line asymmetry fluctuation periods.
This is made possible through combining the results of two-dimensional hydrodynamic simulations with a simplified line radiative transfer model.
The applied grid-based hydrodynamic model allows us to follow the evolution of gas discs over a time scale of 1.2 years (21000 orbits), that has not been previously matched in the literature.

In our model, the gas discs originate from a disrupting asteroid in an eccentric orbit at the sublimation radius of the WD.
Synthetic Ca~II lines are then calculated using an analytic simplified line radiative transfer model.
We give a detailed analysis of the morphology of the lines and investigate the persistence and precession time scales of the discs.
We show that both the DZ nature of the WDs and the asymmetry of the Ca~II lines can persist for over 1.2 years and that our model can replicate the observed precession periods and line morphologies.

This paper is structured as follows.
Section~\ref{sec:modell} describes the numerical models used in this study and the examined parameter space. 
Our results are presented in Sect.~\ref{sec:res}, with each subsection addressing the effect of a single simulation parameter.
Section~\ref{sec:discssion} compares our results to the observed systems.
Finally, our conclusions are given in Sect.~\ref{sec:conc}.

\section{Numerical methods}
\label{sec:modell}

\subsection{Hydrodynamic simulations}
\label{subsec:hidro}

In the hydrodynamic simulations, an asteroid in an eccentric orbit loses material as its orbital distance falls below the sublimation radius of the WD.
The grid-based, two-dimensional hydrodynamic code GFARGO2 is used to describe the gas dynamics.
GFARGO2 is a GPU-based derivative of FARGO \citep{masset00}, further developed in \citet{regaly20} and \citet{regalyetal21}.

\begin{figure}[ht!]
    \centering
    \includegraphics[width=\columnwidth]{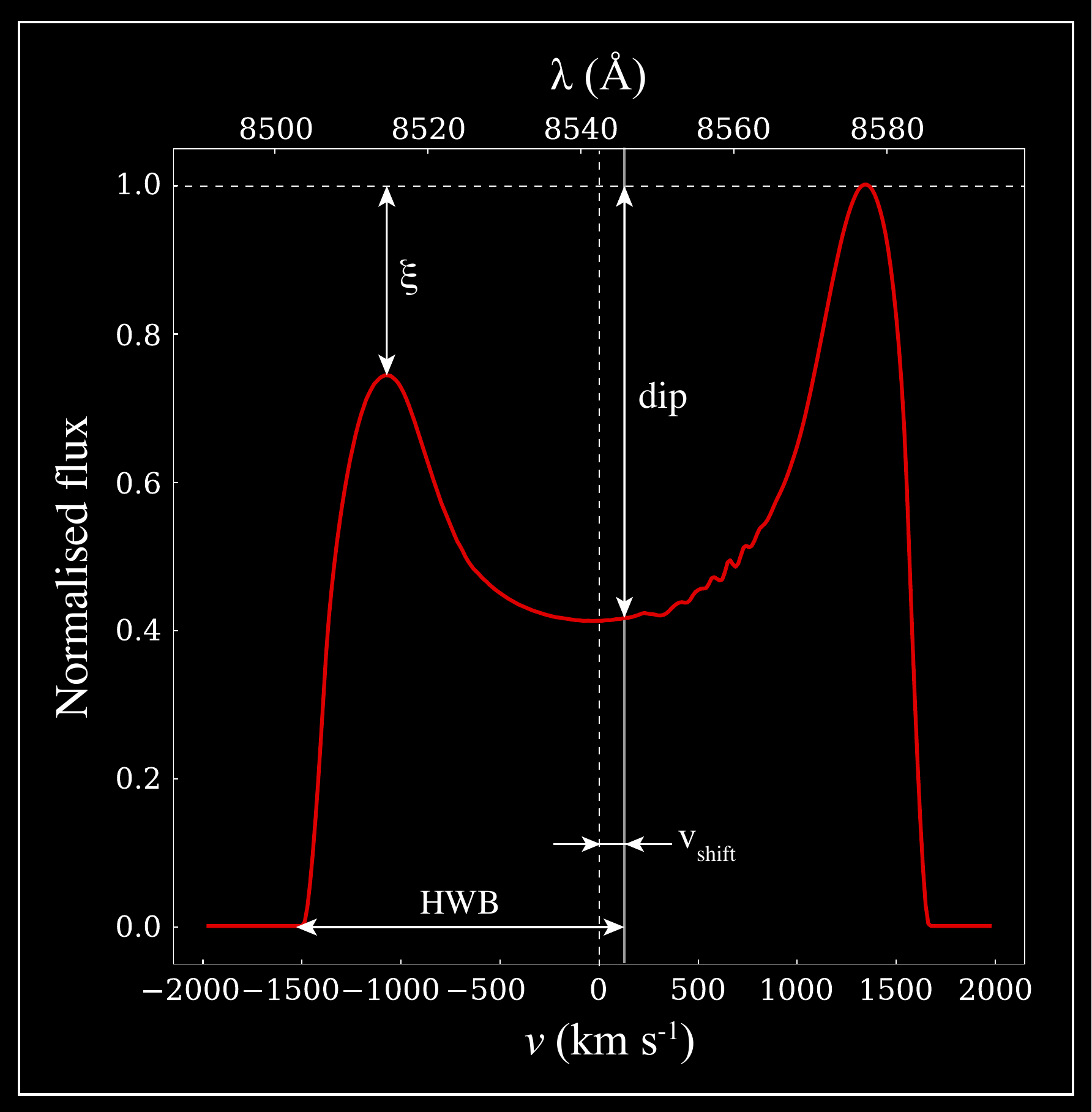}
    \caption{
    Summary of the parameters describing line morphology: asymmetry $\xi$, central dip, shift of the line centre in velocity space ($v_{\mathrm{shift}}$) and half width at base (HWB).
    }
    \label{fig:lineparams}
\end{figure}

This study focuses solely on gas disc evolution.
The hydrodynamic differential equations used are the continuity equation, describing mass conservation, and the Navier-Stokes equation, describing momentum conservation.
\begin{equation}
    \frac{\partial \Sigma}{\partial t} + \nabla(\Sigma \textbf{v}) = \dot{\Sigma}_{\mathrm{a}},
    \label{eq:continuity}
\end{equation}
\begin{equation}
    \frac{\partial \textbf{v}}{\partial t} + (\textbf{v} \nabla) \textbf{v} = - \frac{1}{\Sigma} \nabla P + \nabla \textbf{T} - \nabla \Phi + \dot{ ( \textbf{v}_{\mathrm{a}} \Sigma_{\mathrm{a}} ) },
    \label{eq:navier}
\end{equation}
where $\Sigma$ and $\textbf{v}$ are the surface density and the velocity vector of the gas, while $\Sigma_{\mathrm{a}}$ is the density added during the disruption of the asteroid and $\textbf{v}_{\mathrm{a}}$ is the velocity of the asteroid. 

$\Sigma_{\mathrm{a}}$ is calculated at each time step of the hydrodynamic simulations in a way where material is added to the gas disc based on the mass loss rate of the asteroid derived from a disruption time scale, $\tau_{\mathrm{dis}}$, see later.
The new material is distributed in a two-dimensional Gaussian distribution.
The half width at half maximum of the distribution is calculated as $(r_{\mathrm{a}}h)/2$, where $r_{\mathrm{a}}$ is the radial distance of the asteroid from the star and $h$ is the aspect ratio of the disc.
The density of the new matter is then added to the disc density (see Eq.~\ref{eq:continuity}), and its momentum is calculated from the asteroid's momentum (see Eq.~\ref{eq:navier}) at each time step.
Over time, due to Keplerian shear and gas viscosity, the matter spreads out to form an extended disc.
The process of disc formation is discussed in detail in Sect.~\ref{subsec:early}.

Assuming local thermodynamic equilibrium (LTE), the pressure in the disc can be expressed as $P=c_{\mathrm{s}}^2\Sigma$, where $c_{\mathrm{s}}$ is the local sound speed.
The gravitational potential ($\Phi$) used in the Navier-Stokes equation is:
\begin{equation}
\noindent
    \Phi = 
    -\frac{
    GM_{\mathrm{a}}
    }{
    \sqrt{r^2+r_{\mathrm{a}}^2-2r r_{\mathrm{a}}\cos(\phi-\phi_{\mathrm{a}})+(\epsilon H)^2}}
    -\frac{GM_*}{r}
    -\Phi_{\mathrm{ind}},
\label{eq:gravpot}
\end{equation}
where $G$ is the gravitational constant, $r$, $\phi$, and $r_{\mathrm{a}}$, $\phi_{\mathrm{a}}$ are the radial and azimuthal coordinates of a given grid point and that of the asteroid, respectively.
$M_*$ and $M_{\mathrm{a}}$ denote the mass of the white dwarf and that of the asteroid, respectively. 

The star is positioned at the centre of the grid, which does not coincide with the centre of the inertial frame.
Therefore, the acceleration of the star must be taken into account as an indirect potential term, represented by $\Phi_{\mathrm{ind}}$ in Eq.~(\ref{eq:gravpot}).
Additionally, the gravitational potential of the asteroid must be smoothed to avoid a cell-asteroid distance singularity.
This is achieved through the parameter $\epsilon H$ in Eq.~(\ref{eq:gravpot}), adopting a value of 0.6 based on \citet{kleyetal12}.
The viscous stress tensor is given by
\begin{equation}
    \textbf{T} = \nu \left (\nabla \textbf{v} + \nabla \textbf{v}^T - \frac{2}{3} \nabla \textbf{vI}
    \right ),
\end{equation}
where $\textbf{I}$ is the 2D identity matrix.
Using the \citet{shakurasunaev73} $\alpha$ viscosity parameter, the kinematic viscosity can be expressed by $\nu=\alpha c_{\mathrm{s}} H$.
In the above expression $H$ is the local pressure scale height, $H=hr^{1+\gamma}$, where $h$ is the aspect ratio, and $\gamma$ is the flaring index of the disc.
For of geometrically thin discs, $\gamma=0$, which means that $H/r =h$, where $h$ is a constant.

Numerical algorithms in grid-based hydrodynamic codes are not suitable to describe areas with zero density.
It is therefore necessary to model a background disc in the simulations.
However, since the mass of the background disc is chosen to be one millionth of the mass of the asteroid, its properties do not affect the simulation results.
During the simulations, material from the discs falls onto the star. 
This causes their mass to decrease over time.
If the mass of a disc is less than that of the background, it is considered depleted.

At the inner and outer boundaries, an open boundary condition is applied.
This lets the material flow out from the disc towards the star, but prevents it from flowing into the disc at the inner edge.

\begin{figure*}[ht!]
    \centering
    \includegraphics[width=\textwidth]{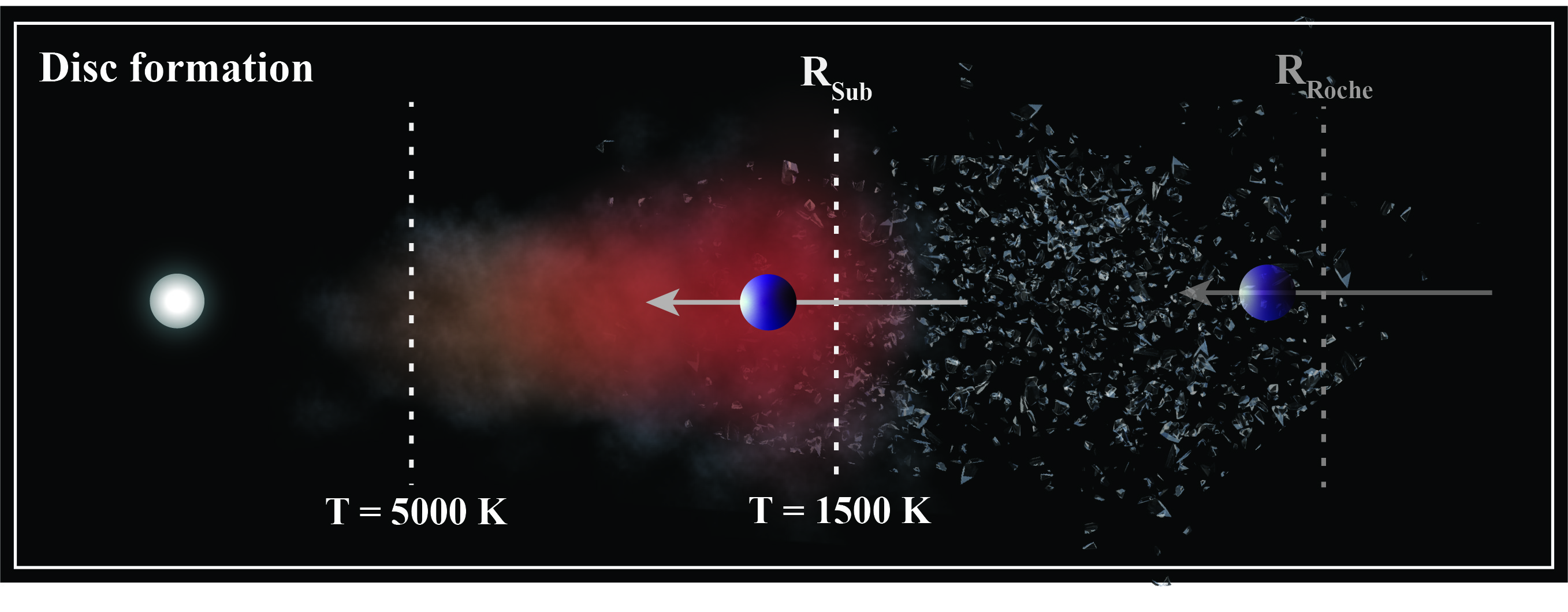}
    \caption{
    The process of disc formation.
    When the asteroid passes under the Roche limit, it loses material, which initially exists as dust until the asteroid passes the sublimation radius ($T=1500~\mathrm{K}$), within which gas is produced.
    The gas then spreads out viscously and forms a disc. 
    The inner limit of Ca~II emission is $T=5000~\mathrm{K}$.
    Note that the figure is not to scale.}
    \label{fig:disc}
\end{figure*}

Given that the mass of the disc is less than that of the embedded asteroids, there isn't enough torque acting on the asteroids to cause them to migrate \citep{verasetal23}.
Therefore, we have disregarded the back-reaction of the gas on the asteroid, so it orbits with a constant semi-major axis.

During the simulations, we monitor the surface density ($\Sigma$), radial velocity ($v_{\mathrm{r}}$), and azimuthal velocity ($v_{\mathrm{t}}$) of the gas.
Analysis of the models also include the examination of the temporal evolution of the asymmetry of the Ca~II line, the disc mass, the total emitted flux and the accretion rate of the WD.

We calculate the Ca~II emission line profiles by implementing a simplified line radiative transfer model, which is detailed in Appendix~\ref{sec:lrt}.
Our assumption that discs are in LTE and are optically thin means that the radial temperature profile in the disc can not be correctly modelled using canonical blackbody radiation.
In order to see the effect of an altered source function of blackbody radiation, we conducted a study where the Planck function is perturbed in either temperature or frequency space.
The results of our analysis are presented in Appendix~\ref{sec:bnu}.
We conclude that perturbing the Planck function in either the temperature or the frequency domain has little to no effect on the Ca~II lines.
As such, we proceed by assuming an unperturbed source function for blackbody radiation.

Line morphology is described by four parameters: degree of asymmetry
($\xi$, the difference in normalised flux between the two peaks of the line); central dip; shift or offset of line centre in velocity space ($v_{\mathrm{shift}}$, the average distance of the two peaks from $0\mathrm{~km~s^{-1}}$); and the half width of the line at base (HWB).
These parameters are visualised in Fig.~\ref{fig:lineparams}.

The temperature profile of the gas is given analytically.
The blackbody temperature profile is defined as
\begin{equation}
    T_{\mathrm{BB}}(r) = \left ( \frac{R_*}{r} \right ) ^{1/2} T_*,
    \label{eq:Tbb}
\end{equation}
where $r$ is the radial distance from the star, while $R_*$ and $T_*$ are the radius and the effective temperature of the WD\footnote{
A number of studies employ the Chiang--Goldreich temperature profile for the discs \citep{jura03, melisetal10}. 
However, this requires the presence of an optically thin, superheated layer of dust above the gaseous component of the disc \citep{chianggoldreich97}. 
This is contrary to the models which are presented here, as all dust is assumed to sublimate above 1500~K, and therefore the Chiang--Goldreich temperature profile is not applicable.}.

Our work assumes that the gas disc is formed by debris falling under the sublimation radius, which can be defined using the above-mentioned temperature profile.
According to \citet{kobayashietal11}, dust is sublimated at 1500~K.

Computational limits based on the Roche radius of the white dwarf and the temperature of the disc are used in the hydrodynamic modelling.
The external computational limit is set at 5/3 times the Roche radius.
The internal computational limits of the simulations in our model correspond to the points at $T=5000~\mathrm{K}$ in accordance with \citet{hartmannetal14}.
Figure~\ref{fig:disc} shows the process of disc formation.

The effects of general relativity are significant at distances less than 1000 times the Schwarzschild radius of the star \citep{sobolenkoetal17}.
However, the inner radius of the gas discs modelled in this paper is over 10,000 times the Schwarzschild radius of the white dwarf.
Therefore, in line with the studies of \citet{manseretal16a, manseretal16b, manseretal19}, the effects of general relativity in the simulations are neglected.

\subsection{Examined parameter space}
\label{subsec:paramspace}

The mass of the white dwarf is set to $M_*=0.7M_\odot$, and its radius to $R_*=4.6\times10^{-5}$~au in all models in accordance with the observations of \citet{gaensickeetal08} and \citet{koesteretal14}
The effective temperature of the star is set to 17000~K, which is consistent with the average temperatures derived by \citet{gaensickeetal07} and \citet{melisetal10}.
The mass of the disrupting asteroid is assumed to be $M_{\mathrm{a}}=10^{-12}$ Earth masses, or $6\times10^{15}$~grams, which corresponds to about the mass of comet 67P/Churyumov--Gerasimenko.

In our model the total mass of the asteroid is converted into gas. Note that the Ca~II content is a free parameter in the simplified, optically thin line radiative transfer model used in this study.  This means that the total emitted Ca~II flux is directly proportional to the Ca~II content. It is important, however, that the line morphology is independent of the injected mass as long as the gas is optically thin. Ca~II content can thus be modified in accordance to the presumed composition of the asteroid, which might be more complex, containing an iron core or even retaining a significant amount of water \citep{malamudperets16, malamudperets17, malamudperets20b, malamudperets20a}.

Assuming a stellar density of $\rho_{*}=3\times10^{6}~\mathrm{g~cm^{-3}}$ \citep{chandrasekhar94}, and an asteroid bulk density of $\rho_{\mathrm{a}}=3~\mathrm{g~cm^{-3}}$, the Roche limit is
\begin{equation}
    R_{\mathrm{Roche}} = R_* \left (3 \frac{\rho_*}{\rho_{\mathrm{a}}} \right ) ^{1/3} \approx 6.6\times10^{-3}~\mathrm{au}.
\end{equation}

The asymmetry and width of the Ca~II emission lines increases with the inclination of the disc.
When the disc is viewed edge-on ($i=90^\circ$), the line-of-sight velocities are maximal.
This study assumes an inclination of $i=45^\circ$ for all models.

Using Eq.~(\ref{eq:Tbb}), the sublimation radius, below which asteroids lose mass, is located at $R_{\mathrm{sub}} = 6\times10^{-3}~\mathrm{au}$.
In this case, given the star's effective temperature of $17000~\mathrm{K}$, the Roche and sublimation radii almost coincide.

\begin{figure}[ht!]
    \centering
    \includegraphics[width=\columnwidth]{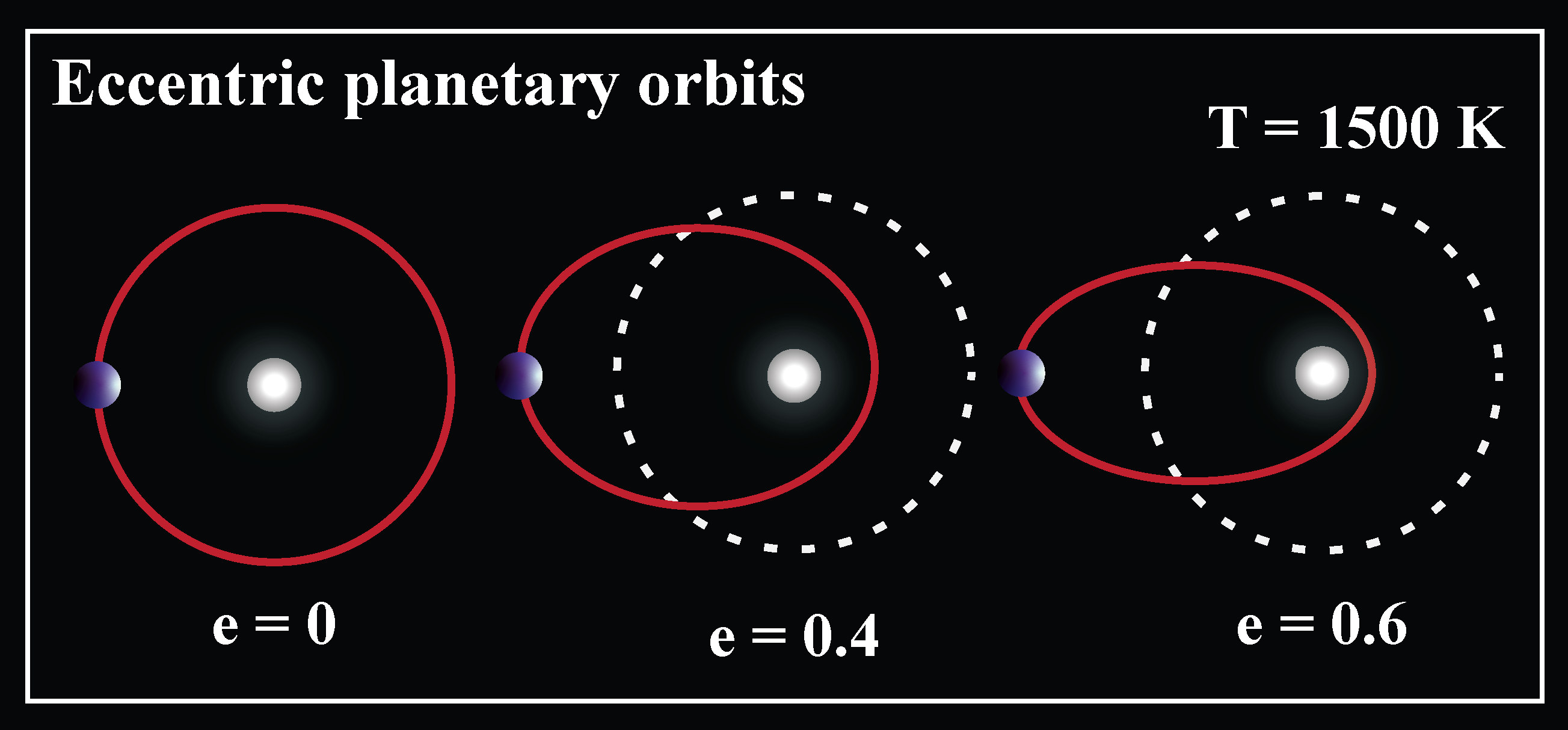}
    \caption{
    The figure shows how asteroid orbits (marked by red lines) are arranged at different eccentricities relative to the sublimation radius (white dashed lines).
    Note that eccentric asteroids spend only part of their orbits below the sublimation radius. 
    }
    \label{fig:eccorbits}
\end{figure}

The semi-major axis of the asteroids always coincides with the sublimation radius, which results in an orbital period of 4.84 hours.
The asteroids are modelled to be moderately eccentric, with $e_{\mathrm{a}}=0.4$ and $0.6$.
The semi-major axis of asteroid is kept constant.
This means that the time the asteroid spends below the sublimation radius varies depending on its eccentricity.
Figure~\ref{fig:eccorbits} illustrates the orbits of such asteroids.

In contrast to the simulations conducted by \citet{trevascusetal21}, the asteroid's disintegration is not assumed to be instantaneous.
The chosen mass loss rate in one set of models ensures that if the asteroid were orbiting on a circular orbit, it would lose all of its mass during one revolution ($\tau_{\mathrm{dis}}$ corresponds to $1~\mathrm{orbit}$).
Other models investigate the effect of disruption that happens on the time scale of the length simulations (i.e. $\tau_{\mathrm{dis}}$ corresponds to $\mathrm{1.2~yr}\approx \mathrm{21000~orbits}$).
It is important to note that hence mass loss only occurs below $R_{\mathrm{sub}}$, disruption on an eccentric orbit takes longer than $\tau_{\mathrm{dis}}$ would dictate.

Viscosity in the SPH simulations of \citet{trevascusetal21} depends on the number of particles modelled: higher particle numbers result in lower viscosity.
However, modelling a large number of particles is computationally expensive, making SPH simulations unsuitable for modelling low viscosities ($\alpha\leq0.001$).
In contrast, the GFARGO2 code used in this work is grid-based and the viscosity is defined in the initial conditions, allowing a wider range to be investigated ($0.001\leq \alpha \leq 0.1$).

The aspect ratio of the disc is set to either $h=0.05$ or $h=0.02$ in our models, the latter of which coincides with the value adopted by \citet{trevascusetal21}.
The discs modelled are assumed to be geometrically thin ($\gamma=0$), in accordance with the assumptions of \citet{jura03} and \citet{veras16}.
The parameters of all models are summarised in Table~\ref{tab:models}.

\begin{table}[h!]
\caption{
    Parameter space of all examined models.}
\centering
    \begin{tabular}{lllllll}
    \hline
    \hline
    \# & $\alpha$ & $\tau_{\mathrm{dis}}$ ($N_{\mathrm{orb}}$) & $e_{\mathrm{a}}$ & $h$ \\ 
    \hline 
    B1 & 0.01  & 1     & 0.4 & 0.05 \\
    B2 & 0.001 & 1     & 0.4 & 0.05 \\
    B3 & 0.05  & 1     & 0.4 & 0.05 \\
    B4 & 0.01  & 1     & 0.4 & 0.02 \\
    B5 & 0.1   & 21 038& 0.4 & 0.05 \\
    B6 & 0.01  & 21 038& 0.4 & 0.05 \\
    B7 & 0.01  & 21 038& 0.6 & 0.05 \\
    \hline
    \end{tabular}
    \tablefoot{
    Columns indicate the model number, viscosity, disruption time (in number of orbits), eccentricity of the asteroid and aspect ratio of the disc.}
    \label{tab:models}
\end{table}

\section{Results}
\label{sec:res}

\subsection{Disc formation}
\label{subsec:early}

\begin{figure}[ht!]
    \centering
    \includegraphics[width=\columnwidth]{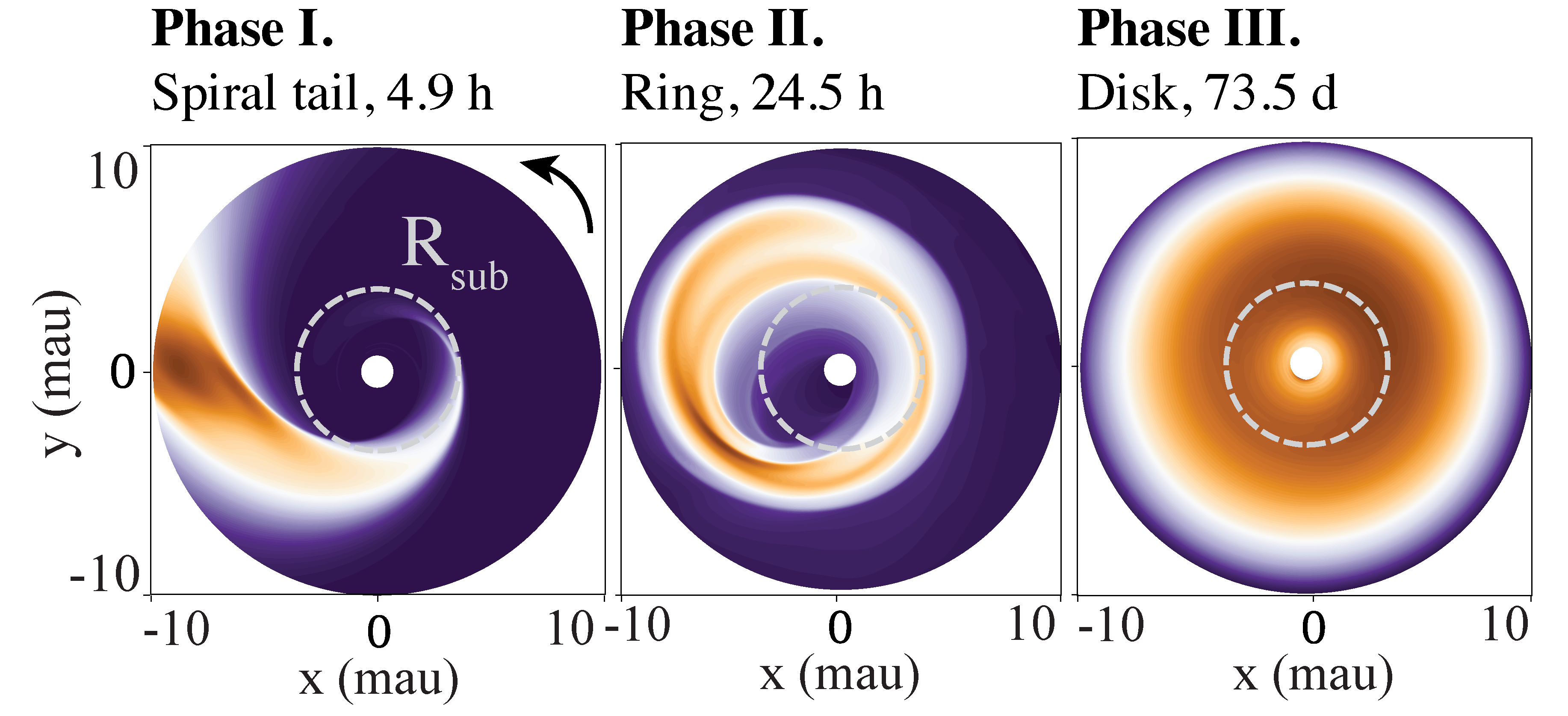}
    \caption{
    Phases of disc formation: snapshots of the density distribution in model B1.
    Grey dashed lines mark the sublimation radius.
    A black arrow shows the orbital direction of the asteroid.
    }
    \label{fig:early}
\end{figure}

\begin{figure*}[ht!]
    \centering
    \includegraphics[width=\textwidth]{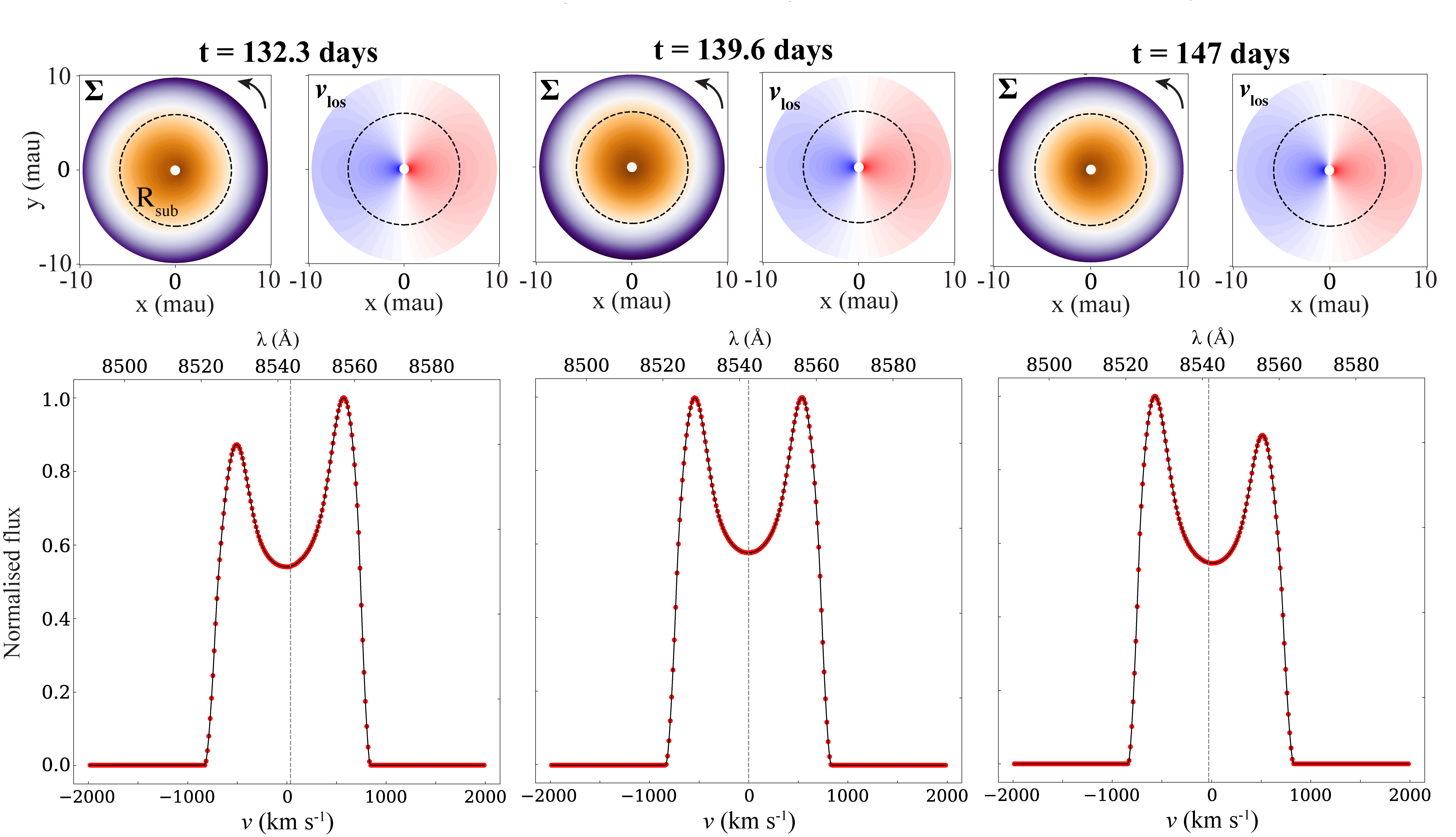}
    \caption{
    Density and velocity distributions ($\Sigma$ and $v_{\mathrm{los}}$, respectively) in model B1 at three points in time (top row) and the corresponding background-normalised Ca~II lines (bottom row).
    The discs are viewed from below.
    The sublimation radius is marked by a black dashed line, and the orbital direction of the planet is shown with a black arrow.
    In the Panels showing the Ca~II lines, vertical dashed lines indicate the centre of the line.
    The retrograde precession of the disc causes a periodic shift in the intensity maximum.
    }
    \label{fig:late}
\end{figure*}

Let us start by making some general statements about the process of disc formation by analysing the model B1 (see parameters in Table~\ref{tab:models}).
Three phases can be identified during disc formation. 
These are shown in Fig.~\ref{fig:early}, where each Panel displays the surface density of the disc at different times.
During the first phase, the gas is pulled apart in a spiral tail along the orbit of the disintegrating asteroid by the Keplerian shear.
In the second phase, the gas forms a ring which begins to precess in the retrograde direction.
At this point, the Ca~II triplet exhibits asymmetry due to the asymmetric density distribution (the effect of the density and velocity distributions is discussed further in Sect.~\ref{subsec:spaces}).
Emphasise that the first and second phases are transient phenomena, meaning that the chance to observe a disc during these phases is low.
However, the second phase can last longer if the viscosity of the disc is lower (see a further discussion in in Sect.~\ref{subsec:viscosity}).
In the third and longest phase, the gas ring viscously spreads to form a disc and starts to accrete onto the white dwarf. 
If the asteroid had an eccentric orbit, the resulting disc will also be eccentric.
Eventually the disc depletes and reaches a mass that is equal to the background disc mass.
Note that there is a continuous transition between each phase.

Changes of the disc density and velocity distributions in phase three of disc formation are shown in Fig.~\ref{fig:late} along with the calculated background-normalised Ca~II line profiles.
The asymmetry of line peaks change on the time scale of the period of the retrograde precession of the disc.
The greatest asymmetry can be seen when viewing the disc from the major axis and the asymmetry is absent when viewing from the minor axis.
The 1.2-year evolution of the Ca~II line profiles in each examined model are shown on spectrograms in Appendix~\ref{app:solit}. 

Steep wings are common to all of the calculated Ca~II lines. This is due to the fact that at the internal computational boundary (i.e. at $T=5000~\mathrm{K}$) Ca~II is converted into Ca~III, as noted by \citet{gaensickeetal06}. Therefore, Ca~II is not observable within this limit.

\begin{figure}[ht!]
    \centering
    \includegraphics[width=0.95\columnwidth]{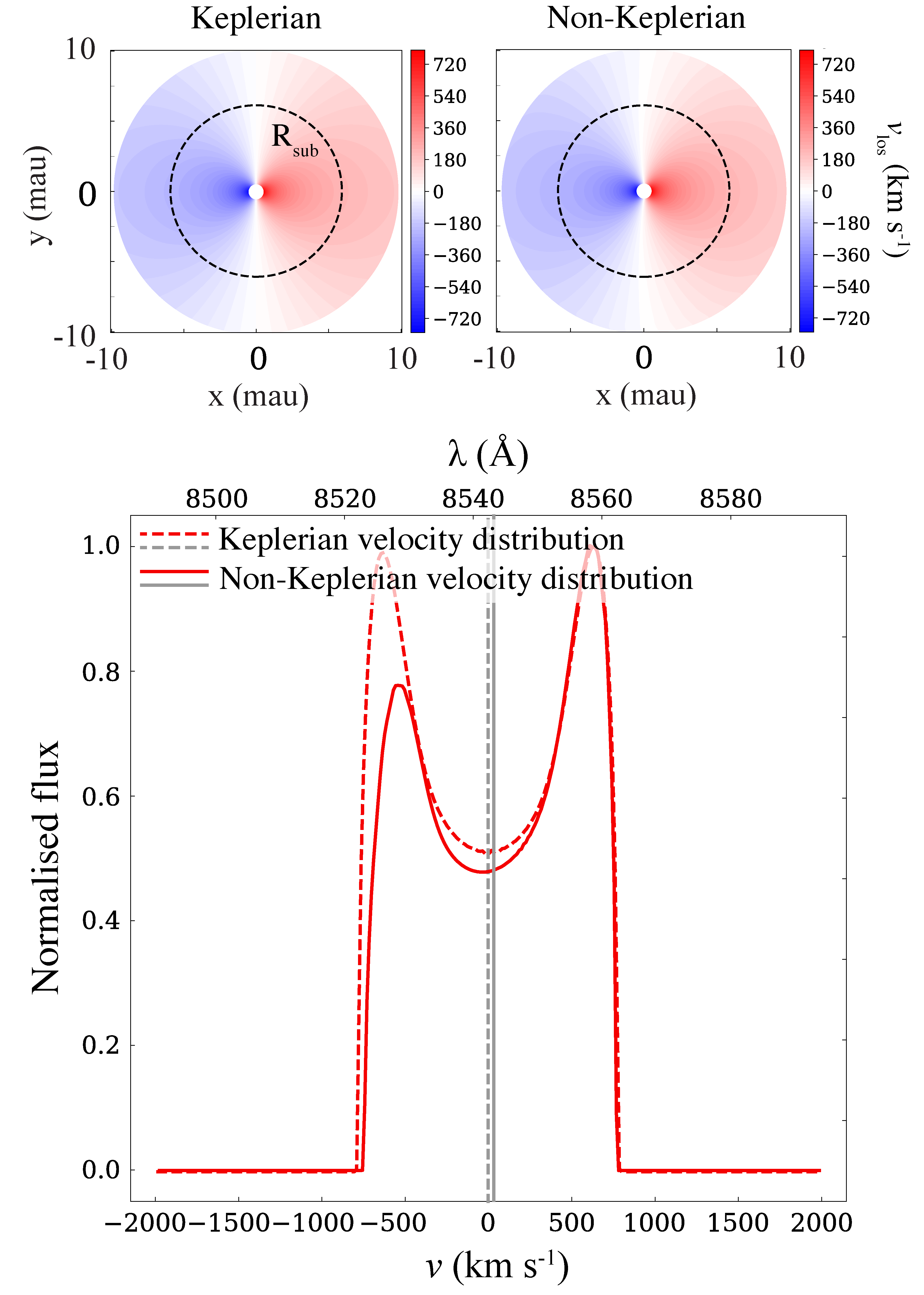}
    \caption{
    The upper Panel displays the line-of-sight velocity distribution in a Keplerian disc, as well as that of model B1 from the hydrodynamic simulations ($t=132.3 \mathrm{days}$).
    The lower Panel shows the calculated Ca~II line profiles with dashed and solid lines in the two cases.
    On the panels showing the velocity distributions, the sublimation radius is indicated by black dashed lines.
    The line centres are marked by vertical grey lines.
    }
    \label{fig:vter}
\end{figure}

\subsection{Origins of the asymmetric Ca~II lines}
\label{subsec:spaces}

In this section we analyse the line profile of model B1 in two cases:
a) when only the density distribution of the hydrodynamic simulations is used when calculating the Ca~II emission, while the velocity distribution is assumed to be Keplerian\footnote{A clump of matter orbiting on a circular orbit can create a Keplerian velocity distribution but a non-Keplerian density distribution, \citealp{redfieldetal17, veraswolszczan19}}; and (b) when both the density and velocity distributions are extracted from the hydrodynamic simulations.

The distribution of the line-of-sight velocity of the gas in the two cases is shown in the top two Panels of Fig.~\ref{fig:vter}.
The left Panel shows the Keplerian velocity distribution, while the right Panel shows the one extracted from the hydrodynamic simulation.
In both cases, the density distribution is the same as the one in Fig.~\ref{fig:late} at $t=132.3 ~\mathrm{days}$.
It is clear that the line-of-sight velocity distribution is symmetric in the Keplerian case.
In contrast, the velocity distribution of the hydrodynamic simulation is perturbed.

\begin{figure}[ht!]
    \centering
    \includegraphics[width=0.9\columnwidth]{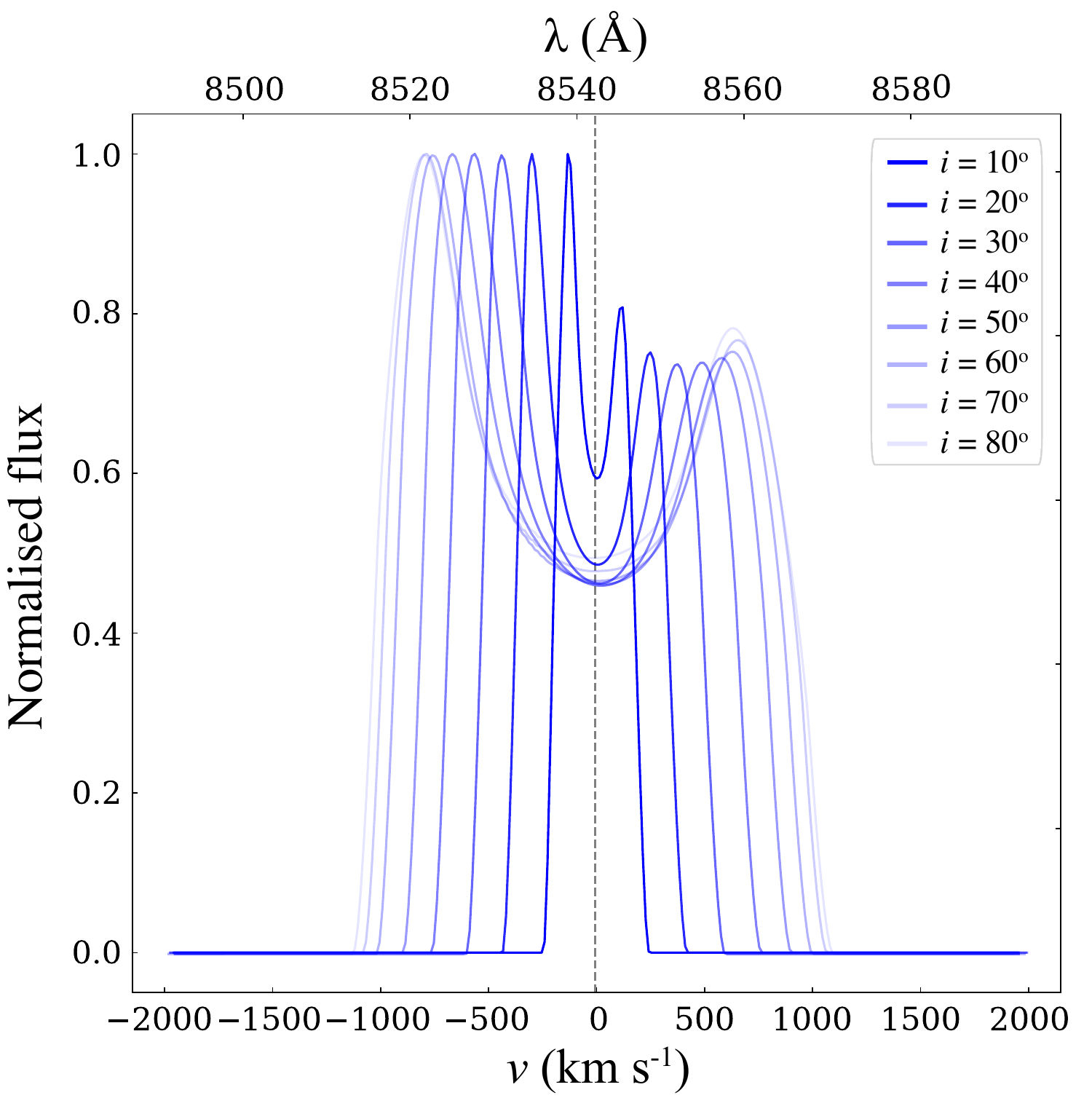}
    \caption{
    Ca~II line profiles for model B1 are plotted with different disc inclinations.
    Fainter lines represent larger inclinations.
    }
    \label{fig:inc}
\end{figure}

The line profiles calculated for each case are shown in the lower Panel of Fig.~\ref{fig:vter}.
Despite the symmetry of the Keplerian case, the Ca~II profile still shows some degree of asymmetry at $\xi=0.06$.
Here, the asymmetric density distribution resulting from the asymmetric material distribution of the eccentric disc is responsible for the asymmetry. 
On the other hand, if the velocity distribution is also non–Keplerian, the asymmetry of the Ca~II line becomes significant, with $\xi=0.22$.

For a Keplerian velocity distribution, the centre of the line is located at $0~\mathrm{km~s^{-1}}$ in all cases.
However, if the velocity distribution is also non–Keplerian, the line centre is shifted (refer to the vertical dashed and solid grey lines in Fig.~\ref{fig:vter}).
If an offset asymmetric line profile is detected, it can be concluded that the gas disc is eccentric and the velocity distribution is asymmetric \citep{regalyetal11, regalyetal14}.
In the model discussed here, the line centre is located at $26.8~\mathrm{km~s^{-1}}$, which corresponds to an offset of 0.74~nm in wavelength space.

The width of the Ca~II lines and location of the line centres are also dependent on the inclination of the gas discs.
Figure~\ref{fig:inc} displays a line profile of model B1 at various angles of inclination ($t=132.3 ~\mathrm{days}$).
The line becomes narrower and less offset as the inclination decreases.
At an inclination of $i=90^\circ$ (i.e. looking at the disc edge-on), the asymmetry is at its maximum.
Additionally, in this case, the line is as wide as possible since the line-of-sight velocities are maximal.
The maximum value of the half-width at bottom (HWB) is $1170~\mathrm{km~s^{-1}}$.
The highest possible offset is $v_{\mathrm{shift}} = 301~\mathrm{km~s^{-1}}$.
Note that in our models the discs are optically thin even when they are viewed edge-on.

\subsection{Disc viscosity}
\label{subsec:viscosity}

This section examines the evolution of discs and the Ca~II line profile while using different $\alpha$ viscosity parameters. 
Values of $\alpha=0.05$, $0.01$ and $0.001$ (models B3, B1 and B2, respectively) are analysed.

We can estimate the depletion timescale of a disc by performing a log-linear fit on the disc mass data, and extrapolating to the point in time when the disc mass is equal to that of the background.
Extrapolation of the data of model B2 yields a depletion timescale of 13.9 years.
However, due to its small viscosity ($\alpha=0.001$), this model evolves very slowly, and fails to develop a disc structure even after 1.2 years. 
Furthermore, by the time the disc structure can fully evolve, the disc circularises.
This means that the Ca~II line profiles will only be asymmetric in the first and second phase of disc formation (see Sect.~\ref{subsec:early}).
This model has not been considered further.

In the $\alpha=0.01$ and $0.05$ models,  
the smaller the viscosity, the weaker the asymmetry:
$\xi=0.13$ in the $\alpha =0.01$ model, and $\xi=0.45$ in the $\alpha =0.05$ model.

The evolution of the disc mass and the flux are correlated:
as the discs start to deplete, the emitted flux also drops by orders of magnitude.
This means that the depletion of the disc also corresponds to the end of the observability for the Ca~II triplet.
As such, measuring the degree of asymmetry and the other line parameters (dip, $v_{\mathrm{shift}}$ and HWB) is only meaningful in the period of best observability, i.e. when the disc is fully formed and its mass is orders of magnitude higher than that of the background disc.

The depletion timescale also depends significantly on the viscosity. 
The mass of the discs decreases more slowly if the viscosity is smaller.
The depletion timescales are estimated to be 1.2 years ($\alpha=0.05$) and 4.3 years ($\alpha=0.01$).

This is also reflected in the total emitted flux:
while the flux in the $\alpha=0.01$ model is saturated at $\simeq10^{26}~\mathrm{erg~s^{-1}~cm^{-2}}$ for the whole runtime, the flux in the $\alpha=0.05$ model shows a constant decrease, and is only $\simeq10^{22}~\mathrm{erg~s^{-1}~cm^{-2}}$ at the end of the simulation.

The same is true for the accretion rate of the WD. 
In model B1, it is constant with a value of $\simeq10^8~\mathrm{g~s^{-1}}$ throughout the 1.2 years of the simulation.
However, in model B3 it drops from $10^8~\mathrm{g~s^{-1}}$ to $10^5~\mathrm{g~s^{-1}}$ in just 0.9 years.

The retrograde disc precession causes a periodic intensity change in the line profile. To determine the precession period $\tau_{\mathrm{p}}$, the Fourier transform of the asymmetry time series is analysed. Again, this is only meaningful to calculate when the disc is fully formed, but still has a significant mass ($10^{-3}~M_{\mathrm{a}}$).

Reducing viscosity not only extends asymmetry and disc lifetime but also significantly influences the precession period of the discs.
The precession periods of the discs analysed in this section are found to be 27.2 and 10.6 days for models B1 and B3, respectively.

For the models considered, the kinematic viscosity and, therefore, the viscous timescale differ.
For geometrically flat discs in the Shakura \& Sunyaev $\alpha$-viscosity prescription, the kinematic viscosity is defined as 
\begin{equation}
    \nu = \alpha h^2 a^{1/2},
    \label{eq:nu}
\end{equation}
which means that the kinematic viscosity in the models considered is $\nu=1.94\times10^{-6}~m^2~s^{-1}$.

The viscous timescale for each disc at the radius corresponding to the semi-major axis of the asteroid, $r_{\mathrm{a}}$, is
\begin{equation}
    t_\nu = \frac{r_{\mathrm{a}}^2}{\nu} \frac{\sqrt{M_*}}{2\pi},
\end{equation}
Table ~\ref{tab:tnu} displays the viscous timescales and precession periods in each model.

\begin{table}[ht!]
    \caption{Viscous timescales in the different models and the precession period of the discs.}
    \centering
    \begin{tabular}{lllll}
        \hline
        \hline
        \# & $\alpha$ & $t_\nu$ (year) & $\tau_{\mathrm{p}}$ (day) \\ \hline
        B1          & 0.01   & 2.4756 & 27.2  \\
        B2\tablefootmark{a}  & 0.001  & 24.755 & -     \\ 
        B3          & 0.05   & 0.4951 & 10.6  \\ 
        \hline
    \end{tabular}
     \tablefoot{ 
    The parameters that differ in each model have also been indicated. \tablefoottext{a}{Precession period can not be determined because a disc structure has not yet developed within the simulation time.}}
    \label{tab:tnu}
\end{table}

\subsection{Disc geometry}
\label{subsec:h}

In this section, model B4 is analysed, where the aspect ratio of the disc is $h=0.02$. 
Its viscosity is $\alpha=0.01$.

The model shows significant line asymmetry, with a maximum of $\xi=0.48$.
The disc precesses at a slower rate than any other model presented in this study, the precession period being 177.4 days.

The decrease in disc mass is slow enough for the Ca~II signal to remain observable after 1.2 years in both models.
The depletion time scales of the disc is found to be  9.5 years.
Comparing these values with model B1 (where $h=0.05$ and the depletion time is 4.3 years), it is evident that the aspect ratio of the discs has a strong effect on the disc lifetime.
The lower aspect ratio of $h=0.002$ results in a disc that persist for about twice as long as their high aspect ratio counterparts.

Regarding the total flux, the model outputs a total flux of  $\simeq5\times10^{26}~\mathrm{erg~s^{-1}~cm^{-2}}$.
The accretion rate of the WD saturates to $\simeq5\times10^{8}~\mathrm{g~s^{-1}}$ and is constant during the simulation.
This accretion rate is different from the B1 model, where $h=0.05$, because in the Shakura \& Sunyaev $\alpha$-prescription, the kinematic viscosity depends on the aspect ratio of the disc (see Eq.~\ref{eq:nu}).

\subsection{Asteroid disruption rate}
\label{subsec:mdot}

In this section, we discuss models where the disintegration of the asteroid occurs in 1.2 years.
Three such models are compared: B5, B6 and B7.

Disc mass is growing throughout all simulations.
This means that the depletion time scale is longer than the simulation time, i.e, all discs persist for at least 1.2 years.
The only time we are able to observe persisting asymmetry is if $\alpha=0.1$.
For the models that use $\alpha=0.01$, the asymmetry declines swiftly, and is completely gone in about 0.5 years.

With regard to the the $\alpha=0.1$ model (B5), the maximum degree of asymmetry is $\xi=0.13$.
The precession period is found to be 126.7 days.
At the end of the simulation, disc mass is saturated to $\simeq10^{-2}~M_{\mathrm{a}}$, the total emitted flux to $\simeq10^{26}~\mathrm{erg~s^{-1}~cm^{-2}}$, and the accretion rate of the WD to $\simeq10^7~\mathrm{g~s^{-1}}$.

Models B6 and B7 were run with different asteroid eccentricities. 
However, they yield the exact same results regarding the monitored parameters.
This means that the initial eccentricity of the asteroid has no effect on the long-term morphology of the lines, or the persistence of the gas disc.

\section{Discussion}
\label{sec:discssion}

\subsection{Comparison to observations}
\label{subsec:vsmeasurements}

According to observations, the accretion rate of the DZ WDs that host gas discs fall in the range of $10^{5}-10^{12}\mathrm{~g~s^{-1}}$ \citep{koesterwilken06,koesteretal14}.
For the models presented here, accretion rates of $10^{5}-5\times10^{8}\mathrm{~g~s^{-1}}$ are derived from the hydrodynamic simulations during the periods of best observability.
It is important to note that the accretion rate is directly proportional to the mass of the asteroid and the above-listed values can be scaled without altering the morphology of the Ca~II lines.

\citet{trevascusetal21} assumes that Ca~II line asymmetry develops in the second phase of disc formation, when material is in a ring around the WD.
While the Ca~II lines are indeed asymmetric in this phase, the gas does not reach the atmosphere of the WD until the third phase, when the disc is fully developed (see Sect.~\ref{subsec:early}).
This means that the DZ nature of the atmosphere and accretion can only be simultaneously observed in the third phase of disc formation.

\begin{table}[h!]
\caption{
    The precession periods and depletion timescales of the models studied, together with the maximum degree of asymmetry, the maximal shift in velocity space, the average dip and the average HWB of the lines.}
\centering
    \begin{tabular}{lrrrrrr}
    \hline
    \hline
    \# & $\tau_{\mathrm{p}}$ (d) & $\tau_{\mathrm{0}}$ (yr) & $\xi$ & $v_{\mathrm{shift}}$ & Dip & HWB \\ 
    \hline
    B1             & 27.2  & 4.3  & 0.13 & 46  & 0.47 & 874 \\ 
    B2\tablefootmark{b}     & -     & 13.9 & -    & 252 & 0.43 & 946 \\
    B3             & 10.6  & 1.3  & 0.45 & 62  & 0.54 & 863 \\
    B4             & 177.4 & 9.5  & 0.30 & 74  & 0.48 & 857 \\
    B5\tablefootmark{a}     & 126.7 & >1.2 & 0.13 & 37  & 0.48 & 852 \\
    B6\tablefootmark{a}     & 147.8 & >1.2 & 0.08 & 50  & 0.45 & 855 \\
    B7\tablefootmark{a}     & 80.6  & >1.2 & 0.07 & 57  & 0.45 & 854 \\
    \hline
    \end{tabular}
     \tablefoot{
     Both $v_{\mathrm{shift}}$ and the HWB are given in units of $\mathrm{km~s^{-1}}$.
    All line parameters are measured when the disc mass is still significant, i.e. $M_{\mathrm{disc}}\geq10^{-3}M_{\mathrm{a}}$.
    \tablefoottext{a}{Discs with $\tau_0>1.2$ are growing in mass throughout the length of the simulations. Determining a depletion timescale would require longer run times. }
    \tablefoottext{b}{Circularisation happens before the gas can evolve into a disc.}
    }
    \label{tab:tprectout}
\end{table}

The observed asymmetric Ca~II lines have a half width at base (HWB) of $130-1150~\mathrm{km~s^{-1}}$, as determined by the Doppler shift of the peaks of the emission lines \citep{dennihyetal20,  gaensickeetal06, gaensickeetal08, gentilefusilloetal21, manseretal16a, melisetal10, melisetal12, melisetal20}.
It is important to note that these observations are degenerate with the inclination of the discs.
In the simulations presented in this work, assuming the disc to be seen as edge-on, the maximum HWB of the Ca~II line is $1170~\mathrm{km~s^{-1}}$.
For WDJ0234-0406 \citep{gentilefusilloetal21}, the HWB of the observed lines is $1150~\mathrm{km~s^{-1}}$, which is only consistent with the simulations if the disc is seen close to edge-on ($i\approx90^\circ$).
The HWB measured in each model, along with other line parameters, are presented in Table~\ref{tab:tprectout}.

As discussed in Sect.~\ref{subsec:spaces}, the centre of the lines shifts when the velocity distribution is asymmetric.
The maximum shift, which occurs for $i=90^\circ$, is $301~\mathrm{km~s^{-1}}$ for all models.
These highest offsets can be observed in model B4, whose aspect ratio is only $0.02$, as opposed to the value of $0.05$ in other models (see Table~\ref{tab:tprectout}).
The above-mentioned offset values are consistent with those of \citet{melisetal10}, who reported line centre displacements of $17-104~\mathrm{km~s^{-1}}$ in three observed systems, and \citet{dennihyetal20}, who measured a shift of up to $260~\mathrm{km~s^{-1}}$ of the Ca~II emission in the HE~1349-2305 system.

The average central dip of the lines ranges from 0.43 to 0.54 (see Table~\ref{tab:tprectout}).
Regarding previous observations, the central dip of 28 (by eye fit) examined Ca~II lines is between 0.13 and 0.8, with about 17 per cent of which lie within the 0.43 to 0.54 range.
It should be noted the values of the central dip shown in Table~\ref{tab:tprectout} are an average over the period of best observability, and can be 0.1 larger or smaller at a given time.
Taking this into account, the models presented here are able to reproduce the central dips seen in 71 per cent of observations.

The central dip of the Ca~II triplet lines is both dependent on the inclination and the optical depth of the disc. 
As shown in Fig.~\ref{fig:inc}, the dip is larger for a larger inclination. 
For sufficiently large inclinations, the disc may become optically thick \citep{hornemarsh86}.
However, this is not the case in the models presented here, as the discs would only become optically thick if their masses were significantly greater (by about 100-1000 times).
Therefore, the observed dip values that differ from those presented here can be explained by a different inclination than that adopted in our models ($45^\circ$).

The line profiles have steep wings in all of our models, and the Ca~II flux is zero in the $T~>~5000~\mathrm{K}$ regions of the discs.
This agrees with the assumption of \citet{gaensickeetal06} that the steep wings are not caused by a complete absence of material, but by additional ionisation of Ca~II into Ca~III.
This is supported by the fact that all WDs with discs have a DZ atmosphere:
the gas does reach the star's surface but is no longer present as Ca~II.

The observed average extent of the gas disc around white dwarfs is $2.35\times 10^{-3} - 5.64\times 10^{-3}$~au \citep{gaensickeetal08, farihietal16}.
For the simulations in this study, the disc extent is between 0.0005 and $1\times 10^{-2}$~au.
These models thus produce discs with larger extents than those observed thus far.

The disruption rate of the asteroid also plays a role in determining the total emitted flux: 
for models where the asteroid disrupts in one orbit, the flux is about five times higher than if it were disintegrating in 1.2 years.
This is a direct consequence of the optically thin LRT model, in which the emitted flux is directly proportional to the disc mass.

\begin{figure}[h!]
    \centering
    \includegraphics[width=0.9\columnwidth]{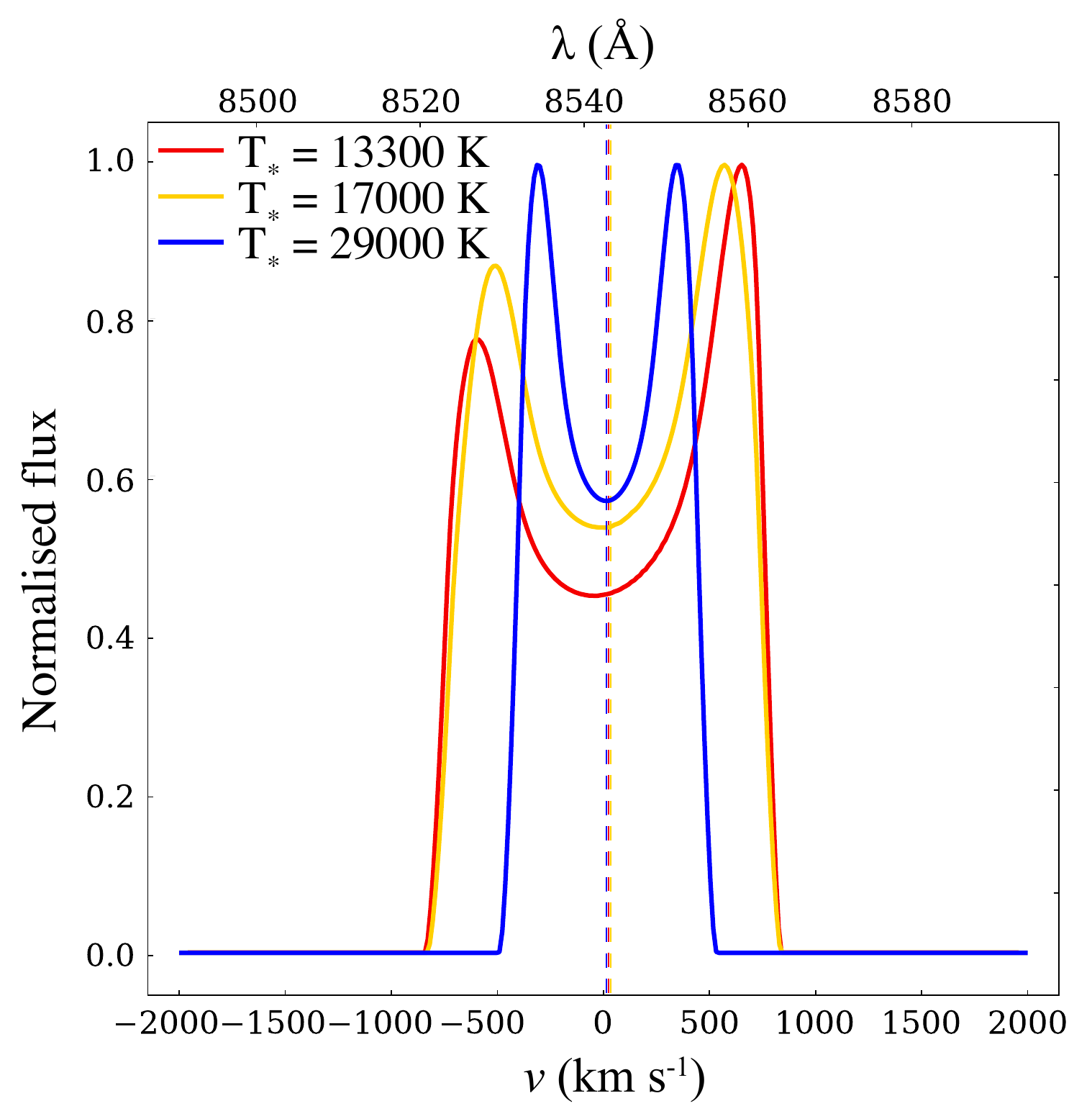}
    \caption{Ca~II line profiles assuming different stellar temperatures in model B1. The snapshots depict the largest possible red-side asymmetry.}
    \label{fig:startemp}
\end{figure}

The models in this work assume a stellar temperature of 17,000~K.
Observations suggest that DZ WDs that host a gas disc have effective temperatures ranging from 13,300~K to 29,000~K \citep{farihietal12, gentilefusilloetal21}.
Figure~\ref{fig:startemp} shows the Ca~II line profile assuming a stellar temperature of 13,300, 17,000 and 29,000~K.
If the stellar temperature is larger, the inner edge of the disc (where $T=5000~\mathrm{K}$) will be farther from the WD.
This implies that the maximal line-of-sight velocities, and thus the HWB of the Ca~II lines, will also be reduced.
The asymmetry of the lines decreases, and can even disappear completely for substantially large temperatures, while the central dip decreases with stellar temperature.
Note that for cooler WDs, the extent of the gas disc is smaller.
In such a case it is necessary to explain how the debris created by the disrupting asteroid on the Roche radius can reach the sublimation radius.
Conversely, around hot stars, the sublimation radius may extend beyond the Roche limit, causing the dust and gas discs to spatially overlap.

\begin{figure}[ht!]
    \centering
    \includegraphics[width=\columnwidth]{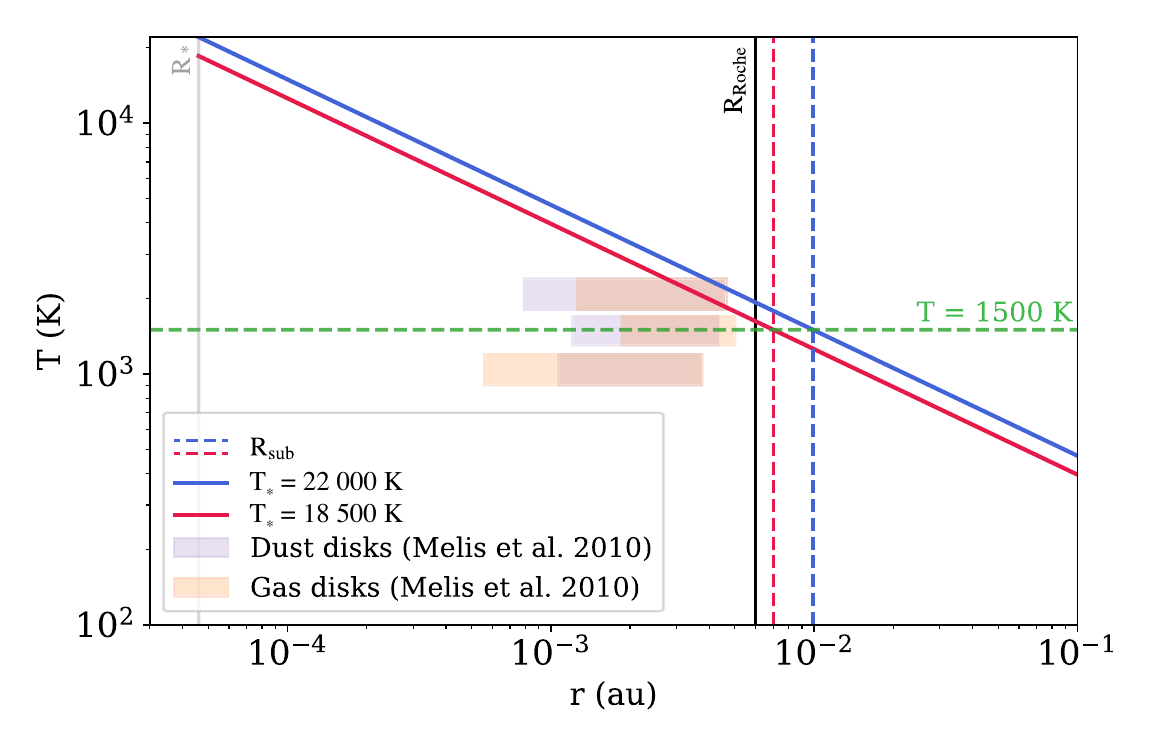}
    \caption{
    The spatial extent of three gas and dust discs that are detected by \citet{melisetal10}.
    The dust discs are marked by grey rectangles, while the gas discs are marked by orange rectangles.
    To improve visibility, the discs for each system are offset vertically.
    For the two stellar effective temperatures used in the \citet{melisetal10} study, namely 18500~K (red lines) and 22000~K (blue lines), the blackbody temperature profile (light red and blue lines) is shown.
    A horizontal green line indicates the sublimation temperature (1500~K).
    The figure displays the sublimation radius for each configuration with vertical dashed lines.
    The dashed grey line indicates the radius of the WD, while the solid black line represents the Roche limit.
    }
    \label{fig:melis}
\end{figure}

Figure~\ref{fig:melis} shows the spatial arrangement of three dust and gas discs presented in \citet{melisetal10}, the extents of which are very similar.
Additionally, the temperature profile investigated in this study is plotted, assuming two stellar temperatures, their corresponding sublimation radii (where $T=1500~K$), and the location of the Roche limit.
Compared to our models, the dust discs appear to be inside the sublimation radius.
The dichotomy presented in Fig.~\ref{fig:melis} can be resolved if the inclination of the gas disc differs significantly from those obtained by \citet{melisetal10}.

The asymmetry of the Ca~II lines has been observed for over twelve years \citep{manseretal16a}.
To maintain this asymmetry for such a long period, it is necessary for the disc to also persist.
According to our simulations, this is possible in two cases:
either when the viscosity is $\alpha \leq 0.01$ and the asteroid disrupts in one orbit, or if $\alpha=0.1$, but the disruption of the asteroid is prolonged to 1.2 years.
Table~\ref{tab:tprectout} summarises the depletion timescale $\tau_{\mathrm{0}}$ for each model, obtained by fitting a log-linear function to the temporal evolution of the disc mass.

The fading of the Ca~II lines has only been observed in the SDSS J1617+1620 system \citep{wilsonetal14}.
The flux increased significantly in 2008 and then showed a steady decrease until 2014 when it disappeared completely.
Similarly, \citet{gentilefusilloetal21} showed a 50 per cent decrease in the flux of Ca~II lines over a three-year period. 
These observations support the idea that disc depletion can occur on a timescale of a few years, resulting in a decrease in the flux of Ca~II lines below what is observable.
Extrapolating the simulations of the first 1.2 years shown in this study, the disc in models B1, B3 and B4 are estimated to deplete in 4.3 and 1.3, and 9.5 years, respectively (see Table~\ref{tab:tprectout}).
These models, therefore, show a rough agreement with the two observations mentioned above.

Based on the periodic fluctuation in the asymmetry of the Ca~II lines, previous research has shown that the precession timescales of the observed gas discs range from a few hours (123 minutes, \citealp{manseretal19}), to a few months (120 days, \citealp{dennihyetal20}), and even several years (1.4-37 years, \citealp{dennihyetal18, manseretal16a, fortinarchambaultetal20, goksuetal23, hartmannetal16}).
The precession period of the models analysed in this paper ranges from 27.2 days to 177.4 days (all precession periods are listed in Table~\ref{tab:tprectout}).
The precession periods of a few months found by \citet{dennihyetal18, dennihyetal20} can thus be explained by these models.
However, they cannot account for the two-hour precession periods measured by \citet{manseretal19} or the years- and decades-long precession periods found by \citet{hartmannetal16, manseretal16a} and \citet{goksuetal23}.
It should be noted that poor temporal resolution of measurements may result in precession periods appearing to be several years or decades long. 
Many of the above-mentioned studies had a time gap of months, and in some cases, a year or two, between two measurements. 
Therefore, if the discs precess at a rate faster than this, the measurements will not be able to detect the rapid variation.

According to the findings presented in Sect.~\ref{subsec:viscosity}, a viscosity of $\alpha\leq0.01$ is required to observe asymmetry for a minimum of 1.2 years.
This is inconsistent with the assumption of \citet{trevascusetal21}, where $\alpha=0.05$.
However, \citet{trevascusetal21} only followed the evolution of the discs for 8.61 days, so no claims can be made about the persistence of the asymmetry over time.
In our case, model B2 (where $\alpha=0.001$) suggests that the viscosity cannot be reduced to any arbitrary value.
The lower the viscosity, the longer the disc remains in the ring phase, however, if the viscosity is too low, the disc will circularise before its material reaches the atmosphere of the WD.
Therefore, it is not possible to simultaneously explain accretion and Ca~II line asymmetry in this scenario.

In the work of \citet{trevascusetal21}, the asteroid has a mass of $1.6\times 10^{-6}$ Earth masses (0.1 times the mass of Ceres).
In this work, the asteroid's mass is assumed to be $10^{-12}$ Earth masses, and the disc is assumed to be optically thin in all simulations.
If the asteroid's mass were to be reduced, the absolute flux of the Ca~II lines would decrease proportionally.
However, the mass of the asteroid does not affect the strength or timescale of the asymmetry as long as the disc is optically thin. 
The discs would only become optically thick if the asteroid's mass were increased above $10^{-10}-10^{-9}$ Earth masses.

The models presented in this work demonstrate that the Ca~II emission signal can be detected over multiple precession periods if the asteroid's disruption occurs on a timescale commensurable to the length of the observations.
This is in a rough agreement with the calculations of
\citet{manseretal19}, which show that a rocky body with a mass of $\leq10^{-9}$ Earth mass orbiting at the Roche radius of the WD can survive for 1.5 years before it completely disrupts.
\citet{malamudperets20a, malamudperets20b} also reach a similar
conclusion: the disruption of a body at a WD's Roche limit does not occur in a single orbit.
The latter is consistent with the models presented in this work, since the asteroids with eccentric orbits also cross the sublimation radius several times.

The presence of minor planets orbiting DZ WDs \citep{vanderburgetal15, manseretal19, vanderboschetal20, guidryetal21} supports the validity of the model presented in this paper.
Around DZ WDs, the asteroid and the gas disc, and thus the Ca~II emission, are observed simultaneously.
This implies that asteroid disintegration and gas production are simultaneous, i.e. disruption does not precede gas formation.

The presence of a second or third body also raises an interesting question.
A giant planet can excite the eccentricity of its companion, causing it to enter the Roche or sublimation radius and triggering the formation of dust or gas.
The reason why we do not see these companion giant planets in the DZ WD systems with transiting exoplanets is likely due to the small size of the star.
Indeed, a giant planet orbiting a WD at $6\times 10^{-3}$~au will not transit for $i\geq0.44^\circ$.

\subsection{Caveats and limitations}
\label{subsec:caveats}

The hydrodynamic model presented in this paper only follows the formation of gas discs.
It does not take into account the asteroid's path to its orbit or the process of disintegration due to the significant computational effort required to describe the multi-component dust and gas discs and their interaction.
The presented model cannot therefore explain the existence of systems where only dust discs are observed.
Nevertheless, several possibilities exist to resolve this:
firstly, future measurement tools may be sensitive enough to detect lower-mass gas discs.
Secondly, it is possible that a gas disc existed in these systems at an earlier time, but viscous spreading caused it to move far away from the star, rendering them undetectable by the Ca~II signal.
It is also possible that the depletion of gas discs does not necessarily mean the depletion of dust discs.

If the Roche limit is outside the sublimation radius, the dust produced on the Roche radius can only approach the WD to a limited extent due to its radiative pressure and the Yarkovsky or YORP effect.
To observe only a dust disc, there must be no dust present under the sublimation radius.
Therefore, a mechanism is required to increase the semi-major axis of the dust particles, which can only be achieved by the Yarkovsky effect assuming a retrograde particle rotation.
It is worth noting that the anomalous metal content of WDs can still be observed for up to thousands to even millions of years, even after the discs have been entirely depleted.

The GFARGO2 code calculates the surface density and velocity of discs in two dimensions.
However, to gain a better understanding of the time evolution of gas discs, three-dimensional studies are necessary.
This would enable the modelling of vertical flows in the disc and the three-dimensional nature of WD accretion (funnel flow, as described in \citealp{romanovaetal03}).

In our model, the internal computational boundary is set at the point where the disc temperature reached 5000~K. 
In this case the process of accretion cannot be satisfactorily modelled.

The simulations presented here do not model the stellar magnetic field.
Instead, its effect is parametrised using the Shakura \& Sunyaev $\alpha$-viscosity.
To accurately describe reality, magneto-hydrodynamic simulations would be necessary.
However, their computational complexity exceeds that of the hydrodynamic simulations discussed in this work.

In this investigation, the gas discs around WDs are assumed to be optically thin.
However, there are observations in the literature that suggest discs are optically thick \citep{gaensickeetal06, jura03}.
Their conclusion is supported by the fact that the centre dip of the Ca~II lines is deeper than expected for an optically thin disc \citep{hornemarsh86}.
As the Ca~II triplet is seen in emission, the emitting region must be optically thin.
However, there can still be an underlying, optically thick layer of gas underneath that has a lower temperature compared to the atmosphere.

Our models assume that the discs are in local thermodynamic equilibrium.
Establishing a full non-LTE model means that the level population of the lines must be derived by taking into account that a) excitation temperatures may deviate from gas (kinematic) temperature, and b) level populations may be strongly radiatively self-coupled.
Furthermore, the temperature dependence of the equations of state must also be taken into account.
This requires solving the energy equation in addition to the continuity and Navier-Stokes equations.
The necessary numerical methods for this latter, computationally expensive task have been implemented in GFARGO2 (see \citealp{tarczaynehezetal20}). 
However, including such complexity is beyond the scope of the current paper.

\section{Conclusions}
\label{sec:conc}

This study examines the evolution of gas discs around white dwarfs and the resulting Ca~II emission.
Gas discs are assumed to be formed by the disruption of asteroids in eccentric orbits at the sublimation radius, which is situated below the Roche–radius of the white dwarf.
The long-term observability of the Ca~II line asymmetry and the metal pollution of the white dwarf are monitored simultaneously.

Hydrodynamic simulations in a locally isothermal regime are carried out on a two-dimensional, cylindrically symmetric grid to model the evolution of the density and velocity distribution of gas discs.
The results of the hydrodynamic simulations are used to calculate synthetic Ca~II emission lines with a simplified line radiative transfer model.
The analysis of the results includes the monitoring of the 1.2-year evolution of the disc mass, the degree of the Ca~II asymmetry, the emitted flux, the accretion rate of the WD and the precession period and depletion time scale of the discs. 
The main results of the simulations are summarised below.
\begin{enumerate}

    \item Asteroids disrupting in eccentric orbits produce globally eccentric, retrogradely precessing discs in three phases: 1) spiral tail; 2) ring; and 3) disc.
    The asymmetry of the Ca~II line is explained by the asymmetric density and velocity distribution of the disc.
    Gas eventually falls onto the white dwarf, explaining its anomalous metallicity.
    
    \item Our results show that the asymmetric Ca~II line profile and the anomalous metallicity of the WD can only be observed simultaneously during the third phase of disc formation.
    This finding resolves the discrepancy of previous results which assume that the Ca~II line asymmetry originates in the second phase of disc formation, when the gas is distributed along a ring and does not reach the surface of the WD.
    
    \item The centre of the Ca~II lines is shifted (by up to $301~\mathrm{km~s^{-1}}$) if the disc velocity distribution is non-Keplerian.
    The shift is a direct indicator of disc eccentricity.

    \item For a moderate inclination of $45^\circ$, the asymmetry of the Ca~II triplet lines ranges from $\xi = 0.13$ to $\xi=0.45$. 
    The width of the lines is approximately $836~\mathrm{km~s^{-1}}$.
    The central dip of the lines ranges, on average, from 0.43 to 0.54.
    The magnitude of the asteroid's eccentricity does not affect the morphology of the lines significantly.
    
    \item There are two ways to explain accretion and asymmetry simultaneously on a time scale of at least 1.2 years:
    a) the asteroid disrupts in one orbit, and disc viscosity is $0.001<\alpha<0.05$;
    b) the asteroid disrupts in a time span of 1.2 years, but the disc viscosity high, in the order of $\alpha=0.1$

    \item Disc precession causes a periodic change of the red and blue peaks of the Ca~II lines whose period falls in the range of 27.2 to 177.4 days.
    
\end{enumerate}

In summary, our work suggests that the persistence of Ca~II asymmetry over decades and its periodic change in the peaks can be explained by asteroids on eccentric orbits in two scenarios.
In the first case, the asteroid disrupts on a short timescale (couple of orbits), and the gas has a low viscosity range ($0.001<\alpha<0.05$) to maintain the Ca~II signal for decades.
In the other scenario, the asteroid disrupts on a a timescale of a year, and the viscosity of the gas is required to be high, $\alpha=0.1$.

The precise origin of WD pollution remains a topic of debate. However, a number of models have been proposed which suggest that small bodies may be scattered or evolve into highly eccentric orbits over time (see, for example, \citealp{verasetal20}).
Moreover, a highly eccentric, comet-like body has been observed to orbit DZ WDs \citep{vanderboschetal20}.
It is therefore important to explore this scenario further in our next study, with a model that considers the disruption of bodies on highly eccentric orbits ($e \gtrsim 0.9$).
Our preliminary findings indicate that in this case, the gas discs must experience a depletion between the pericentre passages of the comet in order for the disc to maintain its eccentricity over a year-long timescale.

The density distributions provided by the hydrodynamic simulations can be used to calculate the transit light curves of slowly disintegrating asteroids, which can be compared to the asymmetric transit light curves of observed objects.
To investigate the nature of the transit asymmetry, implementing the effect of the stellar radiation pressure into GFARGO2 appears promising.

\begin{acknowledgements}
    We thank the anonymous referee, whose helpful comments improved the quality of the paper. The authors would like to express their gratitude to J. Vinkó for fruitful discussions on the nature of line radiative transfer in the discs under consideration, which discussions have had a significant impact on the quality of our study. 
    VF thanks for the financial support from the ÚNKP-23-2 New National Excellence Program of the Ministry for Culture and Innovation from the source of the National Research, Development and Innovation Fund.
    VF acknowledges the financial support of the 'SeismoLab' KKP-137523 \'Elvonal grant of the Hungarian Research, Development and Innovation Office (NKFIH) and the undergraduate research assistant program of Konkoly Observatory.
    
\end{acknowledgements}

\bibliographystyle{aa}
\bibliography{wd} 

\begin{appendix}

\section{Simplified line radiative transfer model}
\label{sec:lrt}

The Ca~II emission line profiles are calculated via a numerical implementation of a simplified radiative transfer model.
The density and velocity distributions obtained from the two-dimensional hydrodynamic simulations are used for the calculations.
The line radiative transfer model assumes that the discs are optically thin.

If the medium is both emitting and absorbing along the propagation of radiation $s$, the basic equation for radiative transfer is:
\begin{equation}
	\label{eq:trans-def}
	\frac{dI_{\nu}(s)}{ds}=j_{\nu}(s)-\chi_{\nu}(s)I_{\nu}(s),
\end{equation}
where $ds$ is the infinitesimal path length along the propagation of the radiation, $I_{\nu}(s)$ is the monochromatic intensity at frequency $\nu$, and $\chi_{\nu}(s)$ and $j_{\nu}(s)$ are the absorption and emission coefficients of the medium at the given frequency.
The monochromatic optical depth along the beam of radiation is $\tau_{\nu}(s)=\chi_{\nu}(s)ds$.
As such, that of a medium with a total thickness of $D$ is
\begin{equation}
	\label{eq:optical-depth}
	\tau_{\nu}=\int_{0}^{D}\chi_{\nu}(s)ds.
\end{equation}
With the above, the radiative transport equation becomes
\begin{equation}
    \label{eq:trans-mod}
    \frac{dI_{\nu}(\tau_{\nu})}{d\tau_{\nu}} =
    \frac{j_{\nu}(\tau_{\nu})}{\chi_{\nu}(\tau_{\nu})}-I_{\nu}(\tau_{\nu})
    = S_{\nu}(\tau_{\nu})-I_{\nu}(\tau_{\nu}),
\end{equation}
where $S_{\nu}(\tau_{\nu})$ is the source function of the emission.
The solution of this equation describes the propagation of a radiation with frequency $\nu$ in a medium with an optical depth of $\tau_\nu$.
If $I_\nu(0)$ is defined as the intensity at the source, it is found in the form of
\begin{equation}
	\label{eq:trans-sol}	
	I_{\nu}(\tau_{\nu}) =I_{\nu}(0)e^{-\tau_{\nu}}+
    \int_{0}^{\tau_{\nu}}S_{\nu}(\tau')e^{-(\tau_{\nu}-\tau')}d\tau'.
\end{equation}
The absorption in a medium with a density of $\rho(s)$ can be given as $\chi_{\nu}(s)=\rho(s)\kappa_{\nu}$, where $\kappa_{\nu}$ is the monochromatic mass absorption coefficient of the medium, assumed to be constant along the path of the radiation, i.e. $\kappa_{\nu}(s)=\kappa_{\nu}$. 
Now Eq.~(\ref{eq:optical-depth}) becomes
\begin{equation}
	\label{eq:optical-depth2}
	\tau_{\nu} = \int_{0}^{D}\rho(s)\kappa_{\nu}ds =  \kappa_{\nu}\int_{0}^{D}\rho(s)=\kappa_{\nu}\Sigma \frac{1}{\cos{(i)}},
\end{equation}
where $\Sigma$ is the density integrated along the path of the radiation, interpreted as the surface density of the disc in the 2D model.
The inclination of the disc is denoted by $i$, and $i=90^\circ$ corresponds to an edge-on disc.

Assuming that the investigated medium radiates as a blackbody, its radiation is described by the Planck function,
\begin{equation}
	\label{eq:Planck}
	B_{\nu}(T)=\frac{2h\nu^{3}}{c^{2}}  \left [ \exp \left( {\frac{h\nu}{kT}} \right) -1 \right ]^{-1},
\end{equation}
where $k$ and $h$ are the Boltzmann, and Planck constants, respectively, and $c$ is the speed of light in vacuum.
It should be noted that the blackbody presumption might be questioned in an optically thin disc in the LTE state.
Nevertheless, modifying the Planck function in both temperature and frequency domains results in a line morphology that is nearly indistinguishable from the one obtained using the canonical source function for blackbody radiation. 
For further details, refer to Appendix~\ref{sec:bnu}.
If the medium is assumed to be in thermal equilibrium, the energy that it emits and that it absorbs must be equal.
In this case $dI_\nu(\tau_\nu)/d\tau_\nu=0$, and applying Eq.~(\ref{eq:trans-mod}) we get Kirchhoff's law, 
$S_\nu(\tau_{\nu})=I_{\nu}(\tau_{\nu})=B_{\nu}(T)$.
This means that the source function in Eq.~(\ref{eq:trans-sol}) can be replaced by the Planck function, and as such, the intensity is
\begin{equation}
	\label{eq:trans-sol-const-dens}
	I_{\nu}=I_{\nu}(0)
 \exp(-\kappa_{\nu}\Sigma)
 +B_{\nu}(T)(1-
 \exp(-\kappa_{\nu}\Sigma)).
\end{equation}
This equation describes that the originally emitted intensity is increased due to the emission of the medium
(second term), while it is decreased due to the
absorption of the medium (third term).

To solve Eq.~(\ref{eq:trans-sol-const-dens}), we need to know the monochromatic mass absorption coefficient, $\kappa_\nu$.
For a simple atomic model with two energy levels, the energy of the lower level is  $\varepsilon_{\mathrm{l}}$, and that of the upper level is $\varepsilon_{\mathrm{u}}=\varepsilon_{\mathrm{l}}+h\nu_0$.
This atom can partake in three different radiation processes.
Firstly, it can spontaneously emit a photon with a frequency of $\nu_0$.
The transition probability for spontaneous emission per unit time is given by the Einstein-coefficient $A_{\mathrm{ul}}$.
Next, it can absorb a photon with a frequency of $\nu_0$.
The probability of this process is proportional to the mean intensity at $\nu_0$, $\bar{J}$, defined as
\begin{equation}
	\bar{J}=\frac{1}{4\pi}\int^{\infty}_0 \int_0^{2\pi}\int_0^{\pi}I_\nu\sin (\theta)d\theta d\varphi \phi_\nu d\nu.
	\label{eq:mean_int}
\end{equation}
The probability of the transition per unit time is $B_{\mathrm{lu}}\bar{J}$, where $B_{\mathrm{lu}}$ is another Einstein-coefficient.
The line profile function $\Phi_\nu$ in Eq.~(\ref{eq:mean_int}) describes the relative effectiveness of neighbouring frequencies of $\nu_0$ for causing a transition.
The third possible radiation process is stimulated emission.
In this case, the transition probability is again proportional to the mean intensity and is given by $B_{\mathrm{ul}}\bar{J}$.
The record of the three radiation processes in local thermodynamical equilibrium, assuming that the number of atoms in the lower state and in the upper state are $n_{\mathrm{l}}$ and $n_{\mathrm{u}}$, respectively, is
\begin{equation}
	n_{\mathrm{u}} A_{\mathrm{ul}}=n_{\mathrm{u}} B_{\mathrm{lu}} \bar{J}-n_{\mathrm{l}} B_{\mathrm{ul}}\bar{J}.
	\label{eq:balance}
\end{equation}
The ratio of the number of species in the upper and lower level states can be given by
\begin{equation}
	\frac{n_{\mathrm{u}}}{n_{\mathrm{l}}}=\frac{g_{\mathrm{l}} 
 \exp \left(-\varepsilon_{\mathrm{l}}/kT \right )
 }{g_{\mathrm{u}} 
 \exp \left(-(\varepsilon_{\mathrm{l}}+h\nu_0)/kT \right )
 }=\frac{g_{\mathrm{u}}}{g_{\mathrm{l}}}
 \exp \left ( -\frac{\varepsilon_{\mathrm{u}}-\varepsilon_{\mathrm{l}}}{kT} \right ),
	\label{eq:levelrat}
\end{equation}
where $g_{\mathrm{l}}$ and $g_{\mathrm{u}}$ are the degeneracy of the lower and upper states, respectively.
Using Eqs.~(\ref{eq:balance}) and (\ref{eq:levelrat}), and due to the fact that in local thermodynamical equilibrium $\bar{J}=B_\nu(T)$ must be valid for all temperatures, the Einstein coefficients ($A_{\mathrm{ul}}$, $B_{\mathrm{ul}}$ and $B_{\mathrm{lu}}$) are coupled by the following relations: 
\begin{equation}
	\label{eq:Einstein-rel}
	B_{\mathrm{ul}}=\frac{c^{2}}{2h\nu_{0}^{3}}A_{\mathrm{ul}},\,\, B_{\mathrm{lu}}=\frac{g_{\mathrm{u}}}{g_{\mathrm{l}}}B_{\mathrm{ul}}.
\end{equation}

The energy absorbed in frequency range $d\nu$, solid angle $d\Omega$ unit time $dt$ by a surface element $dA$ is
\begin{equation}
	dE=\frac{h\nu_0}{4\pi}n_{\mathrm{l}}B_{\mathrm{lu}}\Phi_\nu I_\nu dA\,ds\,dt\,d\Omega\,d\nu.
\end{equation}
We can now apply the relation between energy and intensity: $dE=I_\nu dA\,dt\,d\Omega\,d\nu$.
Using the above two relations, the absorption coefficient of the gas is
\begin{equation}
	\chi_{\nu}(s)=\frac{h\nu}{4\pi}n_{\mathrm{l}}B_{\mathrm{lu}}\Phi_{\nu}.
\end{equation}
We have seen, that stimulated emission is proportional to the mean intensity.
Furthermore, we assume that it only affects the photons along the given beam.
In this case we can view stimulated emission as negative absorption.
As a consequence, the absorption coefficient of the gas corrected for stimulated emission is
\begin{equation}
	\chi_{\nu}(s)=\frac{h\nu}{4\pi}(n_{l}B_{lu}-n_{u}B_{ul})\Phi_{\nu}.
\end{equation}
Gas is assumed to be a locally homogeneous medium, which means that $\rho=nm_{\mathrm{mol}}$, where $m_{\mathrm{mol}}$ is the mass density of the ions.
Both are constant along the propagation of the radiation.
Combining all our previous assumptions, the mass absorption coefficient of the gas can be written as
\begin{equation}
	\kappa_{\nu}=\frac{1}{nm_{\mathrm{mol}}}\frac{h\nu}{4\pi}\left(n_{\mathrm{l}}B_{\mathrm{lu}}-n_{\mathrm{u}}B_{\mathrm{ul}}\right)\Phi_{\nu}.
	\label{eq:mass-abs}
\end{equation}

Assuming only thermal excitation, the total number of the molecules, i.e. the number density, is the sum of molecules in the $i$th state,
\begin{equation}
    n = \sum_{\mathrm{i}}n_{\mathrm{i}} = \frac{n_{0}}{g_{0}} \sum_{\mathrm{i}}g_{\mathrm{i}} 
    \exp \left ( -\frac{\varepsilon_{\mathrm{i}}-\varepsilon_{0}}{kT} \right ).
\end{equation}
In the above expression, $n_0$ is the number of ions in the ground state, and $n_{\mathrm{i}}$ is the number of ions in the $i$th excited state, while $g_0$ and $g_{\mathrm{i}}$ are the degeneracies of the ground and excited states, respectively. 
If the energy of the ground state is $\varepsilon_0$, and that of the excited states are $\varepsilon_{\mathrm{i}}$, we can introduce $Q_{\mathrm{T}}$, the partition function, which can be approximated as follows.
\begin{equation}
	Q_{\mathrm{T}}=\sum_{i}g_{\mathrm{i}}
 \exp \left ( -\frac{\varepsilon_{\mathrm{i}}-\varepsilon_{0}}{kT} \right ).
\end{equation}
The level population of the $i^{\mathrm{th}}$ excited state can be expressed by
\begin{equation}
	\label{eq:population-level}
	n_{\mathrm{i}}=n\frac{g_{\mathrm{i}}}{Q_{\mathrm{T}}}
 \exp \left ( -\frac{\varepsilon_{\mathrm{i}}-\varepsilon_{0}}{kT} \right ).
\end{equation}
It is expedient to find an approximate formula for $Q_{\mathrm{T}}$ as a function of temperature $T$ in a polynomial form, given by
\begin{equation}
	\label{eq:QT-approx}
	\ln(Q_{\mathrm{T}}-g_0)=\sum_{\mathrm{i}}a_{\mathrm{i}}\ln\left(\frac{5040}{T}\right)^{\mathrm{i}}.
\end{equation}
The coefficients $a_{\mathrm{i}}$ for the Ca~II ion can be calculated by the transitional parameters given by \citet{bolton70}, and can be found in Table~\ref{tab:QT-pol-ceff}.

\begin{table}
\caption{
        The coefficients of the partition function of the Ca~II ion.}
	\begin{center}
		\begin{tabular}{cccc}
			\hline 
            \hline 
			$a_{\mathrm{i}}$ & value \\
			\hline
			$a_{0}$	& $-1.582112$ 	\\
			$a_{1}$	& $-3.996089$	\\
			$a_{2}$ & $-1.890737$ 	\\
			$a_{3}$ & $-0.539672$ 	\\	
			\hline
		\end{tabular}
        
        \label{tab:QT-pol-ceff}
	\end{center}
\end{table}

Combining Eqs.~(\ref{eq:Einstein-rel}), (\ref{eq:mass-abs}) and (\ref{eq:population-level}), the monochromatic mass absorption coefficient of the gas at temperature $T$ for a given transition at frequency $\nu_0$ can be expressed
as
\begin{equation}
	\kappa_{\nu}=
    \frac{1}{8\pi}
    \frac{1}{Q_{\mathrm{T}}}
    \frac{A_{\mathrm{ul}}g_{\mathrm{u}}}{m_{\mathrm{mol}}}
    \frac{c^2}{\nu_{0}^2} \left[ 
    \exp \left( -\frac{\varepsilon_{\mathrm{l}}}{kT} \right )
    -
    \exp \left( -\frac{\varepsilon_{\mathrm{u}}}{kT} \right )
    \right]\Phi_{\nu}.
	\label{eq:monochr-abs-coeff}
\end{equation}

Regarding the calculation of the intrinsic line profile function $\Phi_{\nu}$, we have to take into consideration the Doppler line broadening caused by the thermal motion of emitting molecules.
The frequency of the emitted radiation of the molecule in its own frame is $\nu_0$, which corresponds to a frequency of $\nu_0\pm\Delta \nu$ for an observer.
The line emitted by a certain amount of gas is broadened without change in the net flux. 
Assuming a non-relativistic line-of-sight velocity ($v_{\mathrm{los}}/c\ll1$), the Doppler shift causes a change of $\nu-\nu_0=\nu_0 (v_{\mathrm{los}}/c)$ in the observed frequency.
The line-of-sight velocity of the gas in the observer's frame, $v_{\mathrm{los}}$, can be calculated from the velocity components obtained from the hydrodynamical simulations:
$v_{\mathrm{los}} = \left [v_{\mathrm{r}} \sin(\phi) + v_{\mathrm{t}} \cos(\phi) \right ] \cos(i)$, or one can assume it to be Keplerian.
The line-of-sight velocity is responsible for the width of the lines, and the position of the line centre.
Assuming the Maxwell speed distribution for the ions, the number of molecules emitting in the frequency range of $\nu_0+d\nu$ can be written as
\begin{equation}
	n(\nu)= \exp \left [-\left(\frac{c}{\nu_0}\right)^2\frac{m_{\mathrm{mol}}}{2kT}(\nu-\nu_0)^2 \right ].
    \label{eq:N_nu}
\end{equation}

\begin{table}
\caption{Transitional parameters of the Ca~II-triplet.}
\centering
    \begin{tabular}{llll}
    \hline
    \hline
    $\lambda$ (\AA)                                & 8498   & \textbf{8542}  & 8662   \\
    \hline
    $\epsilon_{\mathrm{l}}~(\mathrm{cm^{-1}})$     & 13650  & \textbf{13711} & 13650  \\
    $\epsilon_{\mathrm{u}}~(\mathrm{cm^{-1}})$     & 25414  & \textbf{25414} & 25192  \\
    $A_{\mathrm{ul}}~(10^8~\mathrm{s^{-1}})$ & 0.0111 & \textbf{0.099} & 0.106  \\
    $g_{\mathrm{u}}$                               & 4      & \textbf{4}     & 2      \\ 
    \hline
    \end{tabular}
 \tablefoot{All data are acquired from \citet{bolton70}.
The parameters of the line calculated in this study are boldfaced.}
\label{tab:caii}
\end{table}

Since the strength of the emission (i.e. $\Phi_{\nu}$) is proportional to the number of emitting ions, the thermally broadened line profile is described by a Gaussian function, 
\begin{equation}
	\label{eq:line-profile}
	\Phi_{\nu}=\frac{1}{\sigma\sqrt{\pi}}
 \exp \left [ -\left(\frac{\nu-\nu_{0}}{\sigma}\right)^{2} \right ].
\end{equation}
The broadening of the lines is caused both by thermal excitation and the turbulent movement of the ions.
Thermal broadening is denoted by $\sigma_{\mathrm{term}}$ and, from Eq.~(\ref{eq:N_nu}) can be given as
\begin{equation}
	\label{eq:sigma-therm}
	\sigma_{\mathrm{term}}=\frac{\nu_0}{c}\sqrt{\frac{2kT}{m_{\mathrm{mol}}}}.
\end{equation}
The turbulent line broadening, $\sigma_{\mathrm{turb}}$, is
\begin{equation}
    \sigma_{\mathrm{turb}}=\eta \sqrt{\frac{\Gamma k_{\mathrm{B}} T}{m_{\mathrm{Ca}}}},
\end{equation}
where the adiabatic index of the gas is assumed to be $\Gamma=1.4$, and $m_{\mathrm{Ca}}$ is the mass of the Ca atom.
The turbulent velocity is assumed to be $\eta$ times the sound speed in the gas. 
However, in this study we use the Shakura \& Sunyaev $\alpha$-parameter, so the turbulent broadening becomes
\begin{equation}
    \sigma_{\mathrm{turb}}=\alpha \sqrt{\frac{\Gamma m_{\mathrm{mol}}}{m_{\mathrm{Ca}}}} \sigma_{\mathrm{term}}.
\end{equation}
The total line broadening is then $\sigma = \sqrt{\sigma_{\mathrm{term}}^2+\sigma_{\mathrm{turb}}^2}$.

Due to the gas orbiting the central star, the line profile is shifted by $\Delta \nu$ compared to the $\nu_0$ frequency.
Taking into account the Doppler-shift, the line profile at a given point of the disc ($r$, $\phi$) can be given as
\begin{equation}
	\label{eq:line-profile-mod}
	\Phi_{\nu}=\frac{1}{\sigma\sqrt{\pi}}
 \exp \left [ -\left(\frac{\nu-\nu_{0}+\Delta\nu}{\sigma}\right)^{2} \right ].
\end{equation}
To calculate the final line profile one has to integrate the $\Phi_\nu$ profiles in each cell of the hydrodynamic simulation grid.
Taking into consideration the inclination of the disc, the final line profile is
\begin{equation}
    \Phi = \int \Phi_\nu dA \, \cos{(i)},
\end{equation}
where $dA$ is the area of the emitting grid cell.

The data of all three lines of the Ca~II triplet can be found in Table~\ref{tab:caii} with the line relevant for this work highlighted in bold.
In this study, the 8542~\AA\, wavelength Ca~II line is calculated.

\section{Effects of a perturbed source function for blackbody radiation on line morphology}
\label{sec:bnu}

As previously stated in Appendix~\ref{sec:lrt}, the canonical form of the Planck function described by Eq.~\ref{eq:Planck} may not be able describe an optically thin disc in LTE.
We undertook an analysis of two perturbed Planck functions.
In the first scenario, it is subject to a power-law perturbation in temperature space:
\begin{equation}
	\label{eq:Planck_temp}
	B_{\nu}(T)=\frac{2h\nu^{3}}{c^{2}}  \left [ \exp \left( {\frac{h\nu}{kT}} \right) -1 \right ]^{-1} \left(\frac{T}{T_{\mathrm{in}}}\right)^\beta,
\end{equation}
where $T_{\mathrm{in}}$ is the temperature measured at the inner edge of the disc (5000~K).
In the second case, the function is perturbed in the frequency domain as follows.
\begin{equation}
	\label{eq:Planck_nu}
	B_{\nu}(T)=\frac{2h\nu^{3} \nu^*}{c^{2}}  \left [ \exp \left( {\frac{h\nu \nu^*}{kT}} \right) -1 \right ]^{-1},
\end{equation}
\begin{equation}
    \nu^*=\left ( \frac{\nu}{\nu_0}\right )^\alpha, 
    \label{eq:nustar}
\end{equation}
where $\nu_0$ is the centre of the $\lambda=8542~\AA$ line.

\begin{figure}[ht!]
    \centering
    \includegraphics[width=0.9\columnwidth]{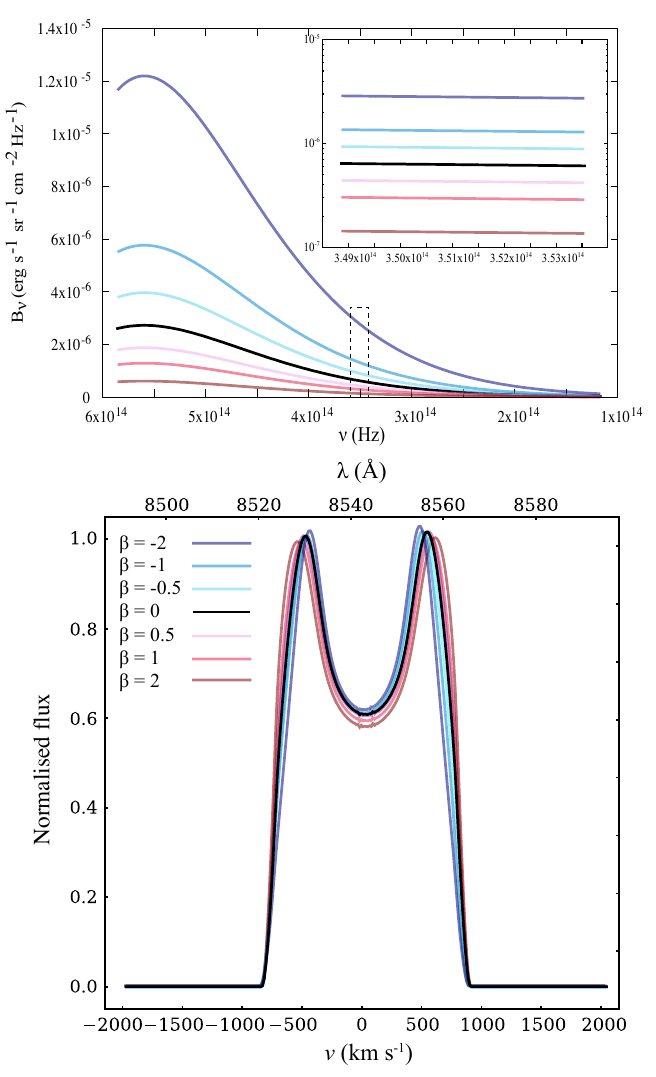}
    \caption{
    Top: Planck functions in the frequency domain. Different colours correspond to different perturbations in the temperature domain (see Eq.~\ref{eq:Planck_temp}). The unperturbed model is denoted by a black line. The inlet shows the close proximity of the Ca~II triplet, marked by a dashed rectangle.
    Bottom: background-normalised $\lambda=8542~\AA$ Ca~II line profiles calculated using the perturbed $B_\nu$ functions.
    }
    \label{fig:bnu-temp}
\end{figure}

The top panel of Fig.~\ref{fig:bnu-temp} illustrates $B_{\nu}$ as a function of frequency, with varying values of $\beta$.
The bottom panel depicts the calculated $\lambda=8542~\AA$ emission line for each case, assuming a Keplerian disc that is unperturbed in both density and velocity space.
As can be observed, the morphology of the background-normalised lines does not exhibit significant variation from the unperturbed case.
However, the absolute flux of the lines undergoes a slight alteration in response to the perturbation on a range from $7\times10^{13}~\mathrm{erg~s^{-1}~cm^{-2}~Hz^{-1}}$ at $\beta=2$ to $2.25\times10^{14}~\mathrm{erg~s^{-1}~cm^{-2}~Hz^{-1}}$ at $\beta=-2$.

\begin{figure}[h!]
    \centering
    \includegraphics[width=1\columnwidth]{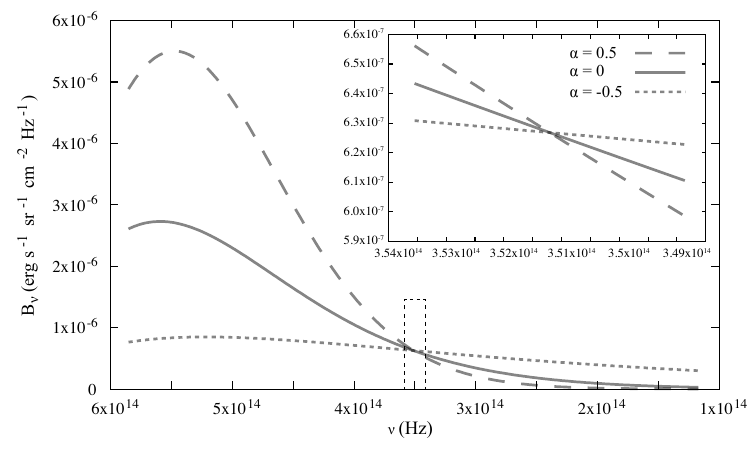}
    \caption{
    Same as the left panel of Fig.~\ref{fig:bnu-temp}, but with the perturbation occurring in frequency space (see Eq.~\ref{eq:Planck_nu}). The unperturbed model is shown by the solid black line. 
    }
    \label{fig:bnu-nu}
\end{figure}

Figure~\ref{fig:bnu-nu} depicts $B_{\nu}$ as a function of frequency, with varying values of $\alpha$. 
The lines perturbed in frequency space are not shown, as they are practically indistinguishable from the unperturbed line profiles, both in terms of morphology and absolute flux.
Furthermore, selecting a value of $\nu_0$ that is either half or twice the initial value (the value corresponding to $\lambda=8542~\AA$) does not result in a notable change in the background-normalised line morphology.

Based on our analysis, neither the temperature nor the frequency space perturbations cause a difference in the relative strength of the Ca~II triplet lines.
This is due to the fact that the applied perturbation on the Planck function results in a very small variation in the close vicinity of the Ca~II triplet.
As a summary, we conclude that the applied perturbations have little to no effect on the line morphologies, and we proceed with assuming an canonical source function for blackbody radiation.

\section{Spectrograms}
\label{app:solit}

\begin{figure*}[h!]
    \centering
    \includegraphics[width=\textwidth]{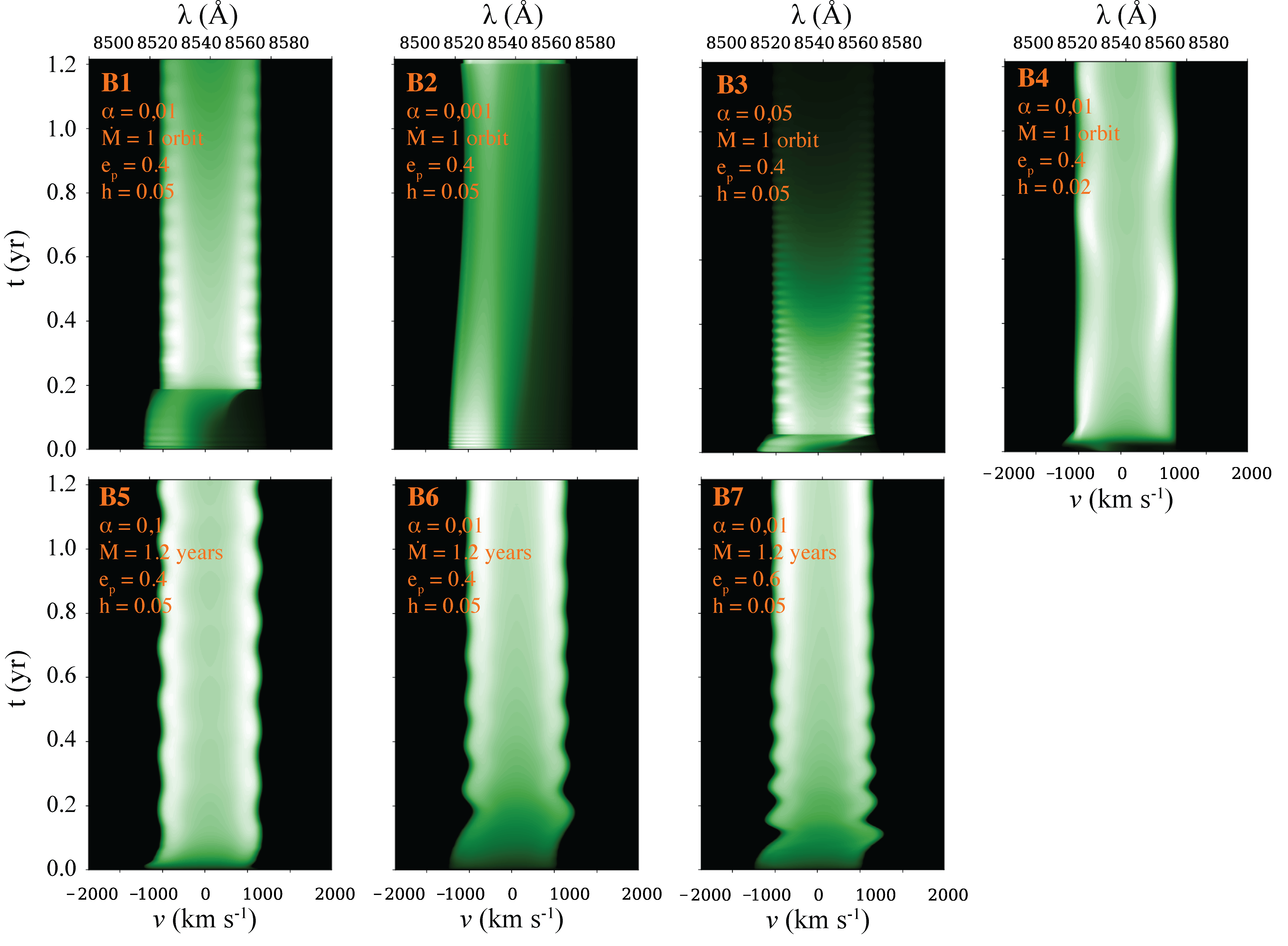}
    \caption{
    Spectrograms of all models. Colours code the absolute flux of the Ca~II lines. 
    }
    \label{fig:visc_solitosis}
\end{figure*}

Figure~\ref{fig:visc_solitosis} displays spectrograms of all models, showing a series of Ca~II line profiles as a function of time in velocity and wavelength space.
The decrease of line intensity can clearly be observed (see, e.g. model B3), which corresponds to the decrease in flux and disc mass and the eventual depletion of the disc.

\end{appendix}

\label{LastPage}
\end{document}